\documentclass[12pt,a4paper]{article}
\pdfoutput=1

\usepackage[utf8]{inputenc}
\usepackage[T1]{fontenc}
\usepackage{ dsfont }
\usepackage{multirow}
\usepackage{cite}
\usepackage{amsmath}
\usepackage{color}
\usepackage{amsfonts}
\usepackage{amssymb}
\usepackage{graphicx}
\usepackage{geometry}
\usepackage{amssymb,epsfig,subfigure}
\usepackage{hyperref}
\usepackage{url}
\usepackage{comment}
\usepackage[font=footnotesize]{caption}
\usepackage{slashed}
\newcommand{\cev}[1]{\reflectbox{\ensuremath{\vec{\reflectbox{\ensuremath{#1}}}}}}
\usepackage{tensor}
\usepackage{cancel}
\makeatletter
\renewcommand\section{\@startsection {section}{1}{\z@}%
                                 {-3.5ex \@plus -1ex \@minus -.2ex}
                                   {2.3ex \@plus.2ex}%
                                   {\normalfont\large\bfseries}}
\renewcommand\subsection{\@startsection{subsection}{2}{\z@}%
                                   {-3.25ex\@plus -1ex \@minus -.2ex}%
                                     {1.5ex \@plus .2ex}%
                                     {\normalfont\bfseries}}
\renewcommand\subsubsection{\@startsection{subsubsection}{3}{\z@}%
                                   {-3.25ex\@plus -1ex \@minus -.2ex}%
                                     {1.5ex \@plus .2ex}%
                                     {\normalfont\itshape}}
\makeatother

\def\pplogo{\vbox{\kern-\headheight\kern -29pt
\halign{##&##\hfil\cr&{\ppnumber}\cr\rule{0pt}{2.5ex}&\ppdate\cr}}}
\makeatletter
\def\ps@firstpage{\ps@empty \def\@oddhead{\hss\pplogo}%
  \let\@evenhead\@oddhead 
}
\def\maketitle{\par
 \begingroup
 \def\thefootnote{\fnsymbol{footnote}}
 \def\@makefnmark{\hbox{$^{\@thefnmark}$\hss}}
 \if@twocolumn
 \twocolumn[\@maketitle]
 \else \newpage
 \global\@topnum\z@ \@maketitle \fi\thispagestyle{firstpage}\@thanks
 \endgroup
 \setcounter{footnote}{0}
 \let\maketitle\relax
 \let\@maketitle\relax
 \gdef\@thanks{}\gdef\@author{}\gdef\@title{}\let\thanks\relax}
\makeatother
\numberwithin{equation}{section}

\newcommand\eea{\end{eqnarray}}
\newcommand\bea{\begin{eqnarray}}

\def\beq{\begin{equation}}
\def\eeq{\end{equation}}

\newcommand{\be}{\begin{equation}}
\newcommand{\ee}{\end{equation}}
\newcommand{\ba}{\begin{align}}
\newcommand{\ea}{\end{align}}
\newcommand{\bg}{\begin{gather}}
\newcommand{\eg}{\end{gather}}
\newcommand{\bseq}{\begin{subequations}}
\newcommand{\eseq}{\end{subequations}}

\newcommand{\p}{P}

\newcommand{\coment}[1]{}

\begin{document}
\setcounter{page}0
\def\ppnumber{\vbox{\baselineskip14pt
}}
\def\ppdate{
} \date{}

\author{Valentin Benedetti$^1$, Lucas Daguerre$^2$\\
[7mm] \\
{\normalsize \it $^1$Centro At\'omico Bariloche and CONICET}\\
{\normalsize \it S.C. de Bariloche, R\'io Negro, R8402AGP, Argentina}\\
{}\\
{\normalsize \it $^2$ Center for Quantum Mathematics and Physics (QMAP)}\\
{\normalsize \it Department of Physics \& Astronomy, University of California, Davis, CA 95616 USA}\\
}

\bigskip
\title{\bf Entanglement entropy of a\\  Rarita-Schwinger field in a sphere\vskip 0.5cm}
\maketitle

\begin{abstract}
We study the universal logarithmic coefficient of the entanglement entropy (EE) in a sphere for free fermionic field theories in a $d=4$ Minkowski spacetime. As a warm-up, we revisit the free massless spin-$1/2$ field case by employing a dimensional reduction to  the $d=2$ half-line and a subsequent numerical real-time computation on a lattice.  Surprisingly, the area coefficient diverges for a radial discretization but is finite for a geometric regularization induced by the mutual information. The resultant universal logarithmic coefficient $-11/90$ is consistent with the literature. For the free massless spin-$3/2$ field, the Rarita-Schwinger field, we also perform a dimensional reduction to the half-line. The reduced Hamiltonian coincides with the spin-$1/2$ one, except for the omission of the lowest total angular momentum modes. This gives a universal logarithmic coefficient of $-71/90$. We discuss the physical interpretation of the universal logarithmic coefficient for free higher spin field theories without a stress-energy tensor.
\end{abstract}
\bigskip\bigskip\bigskip\bigskip\bigskip\bigskip
{\small{\vspace{1.5 cm}\noindent ${}^{1}\,\,$valentin.benedetti@ib.edu.ar\\
${}^{2}\,\,$ldaguerre@ucdavis.edu }}


\newpage

\tableofcontents

\vskip 1cm

\newpage
\section{Introduction} 
\label{sec:intro}
The entanglement of vacuum fluctuations is an ubiquitous property of relativistic quantum fields, as implied by the Reeh-Schlieder theorem \cite{Witten:2018zxz}. Moreover, in recent years, it has proven to be relevant to the study of quantum field theory  and quantum gravity \cite{Casini:2022rlv,Faulkner:2022mlp}. Specifically, given a vacuum state and a von Neumann algebra of observable operators localized inside a region, the entanglement entropy (EE) can be used to provide a measure of entanglement between such region and its complement. For spherical regions, the assignation of an algebra in the continuum is unique \cite{Casini:2020rgj}. Further, in any conformal field theory (CFT), the  structure of the corresponding EE for a sphere of radius $R$ embedded in a $d=4$ Minkowski spacetime takes the form
\begin{equation}
    S(R)=c_2 \,\frac{R^2}{\epsilon^2}+c_{\text{log}}\log\left(\frac{R}{\epsilon}\right)+c_0,
    \label{eq:EntropyCFT}
\end{equation}
where $\epsilon$ represents any given short-distance cutoff introduced to regulate the UV divergences of the EE. The area coefficient $c_2$ is regularization dependent and might even diverge for a given scheme \cite{Riera:2006vj}. Also, the constant coefficient $c_0$ is cutoff dependent. On the other hand,  the logarithmic coefficient $c_{\text{log}}$ is a universal quantity of the theory because it is well defined in the continuum limit. Remarkably, this universal coefficient has been used to provide an entropic proof of the irreversibility A-theorem \cite{Casini:2017vbe,Casini:2023kyj}, so it can be regarded as an effective measure of the degrees of freedom in the underlying theory at a given energy scale. Irreversibility theorems associated to the subleading coefficients have also been established \cite{Casini:2014yca,Casini:2015ffa,Casini:2017vbe,Daguerre:2022uxt}. Indeed, the logarithmic coefficient is proportional to the type-A trace anomaly, which can be extracted from the coefficient of the expectation value of the trace of the stress–energy tensor proportional to the Euler density in curved space-time \cite{Deser:1993yx}.  This has been proven using holography \cite{Solodukhin:2008dh}, as well as for general CFTs with a well-defined stress–energy tensor using a conformal map to de Sitter spacetime \cite{Casini:2011kv}.

The logarithmic coefficients $c_{\text{log}}(h)$ for all physical massless fields with spin $h \leq 1$  were computed from the type A trace-anomaly using euclidean methods \cite{Solodukhin:2011gn,Birrell:1982ix}, yielding 
\begin{equation}
    c_{\text{log}}(0)=-\frac{1}{90}\,,\quad c_{\text{log}}(1/2)=-\frac{11}{90}\,,\quad c_{\text{log}}(1)=-\frac{31}{45}\,.
    \label{eq:clogLowSpinLess1}
\end{equation}
 These coefficients have also been  computed by other means both analytically \cite{Casini:2010kt,Dowker:2010bu,Huang:2014pfa,Huerta:2022tpq} and numerically \cite{Lohmayer:2009sq,Casini:2015dsg}.  For the  cases of Klein-Gordon scalar and Dirac fermion fields all computations coincide. However, for the free Maxwell field, explicit field theoretic calculations performed in \cite{Casini:2015dsg}, as well as thermodynamic computations in de Sitter spacetime \cite{Dowker:2010bu}, reveal a different result
\begin{equation}
     \tilde{c}_{\text{log}}(1)=-\frac{16}{45}\,.
    \label{eq:clogLowSpinLessdif}
\end{equation}
The reason behind this mismatch is that the result for the entropy depends on the details of the operators inserted at the boundary of the entangling region, together with the possible non-local correlations induced by them. This ambiguity appears in all theories that are not complete, in the sense that they exhibit non-local sectors where the  presence of these operators at the boundary can modify the expected anomaly result \cite{Casini:2019nmu}. For example, in the Maxwell case, the difference between these two computations is that (\ref{eq:clogLowSpinLessdif}) corresponds  to a free Maxwell theory, while (\ref{eq:clogLowSpinLess1}) corresponds to an effective Maxwell theory in the IR where interactions produced by heavy electric and magnetic charges appearing in the UV completion were integrated out. This coupling to charges in the UV can also be interpreted as the physical origin of the ``edge modes'' proposed in \cite{Huang:2014pfa,Donnelly:2014fua,Donnelly:2015hxa,Soni:2016ogt} for the IR theory.

For theories with spin $h\geq 3/2$ there seems to be a more  severe problem, as the Weinberg-Witten theorem \cite{Weinberg:1980kq} forbids the existence of a well-defined stress-energy tensor. This makes the notion of trace anomaly unclear from the field theory point of view \cite{Birrell:1982ix}.  However, in such cases, the logarithmic coefficient can still be computed without invoking the trace anomaly. Such calculations have been performed for the spin 2 case in \cite{Benedetti:2019uej, David:2020mls}. Even more, it was conjectured \cite{Benedetti:2019uej, Dowker:2019zva} and proved using the replica trick \cite{David:2020mls} that the logarithmic coefficients for any field of integer spin $h$ are given by
\begin{equation}
    c_{\text{log}}^{\text{bos}}(h)=-g(h)\frac{1+15h^2}{90}, \:\:\:\:\:\:\:\:\: g(h)=\begin{cases}
    1\:\:\:\:\:\:h=0\\
    2\:\:\:\:\:\:h>0
    \end{cases}.
\end{equation}
The result for half-integer spin $h$ fields was also proposed by \cite{Dowker:2019zva}
\begin{equation}
    c_{\text{log}}^{\text{fer}}(h)=-\frac{7+60h^2}{180}.
    \label{eq:clogFerm}
\end{equation}
By studying the universal logarithmic coefficients in the EE for free higher spin theories, we aim to gain more insight regarding its physical interpretation.  Following this line, we begin by performing the explicit real-time computation of the logarithmic coefficient associated with a massless spin-$3/2$ field described by the free Rarita-Schwinger field \cite{Rarita:1941mf,freedman2012,Weinberg:2000cr}. We do so by working with the approach developed in \cite{Srednicki:1993im,Lohmayer:2009sq,Casini:2015dsg,Benedetti:2019uej}. The procedure consists of using a harmonic mode decomposition on the transverse sphere $S^2$, and a subsequent dimensional reduction that produces a $d=2$ effective radial Hamiltonian. The former is made of a tower of modes labeled by the total angular momentum eigenvalues $j$. The EE of each mode can be computed numerically on a lattice and the final result follows from a careful addition over modes. Surprisingly, for spin $1/2$ fields, this calculation has not been done in the context of an EE computation. We do provide such computation. For spin $3/2$ fields, we prove that the effective radial Hamiltonian obtained matches with the spin $1/2$ result without the lowest total angular momentum mode $j=1/2$. We will demonstrate  that this result guarantees that (\ref{eq:clogFerm}) holds for $h=3/2$. 

The structure of the article is the following: in Section \ref{sec:Dirac}, we dimensionally reduce a massless Dirac field in $d=4$ in a sphere using spinor spherical harmonics. We use this result to compute numerically the EE in a sphere. The radial discretization \cite{Srednicki:1993im} gives a divergent area coefficient, so the logarithmic coefficient is obtained from a regularized EE defined in terms of the mutual information \cite{Casini:2015woa}. In Section \ref{sec:RaritaSTheory}, we review the free Rarita-Schwinger theory in $d=4$. We mainly work with the gauge invariant phase space generated by the field strengths that can be defined from the usual gauge-dependent vector-spinor fields. Subsequently, in Section \ref{sec:RaritaS}, we decompose its massless Hamiltonian using spinor-vector spherical harmonics. We employ the same ideas used in the spin $1/2$  case to conclude that the logarithmic coefficient is $-71/90$. Finally, Section \ref{sec:concl} is devoted to a discussion regarding the existence of ambiguities in the logarithmic coefficient of free higher spin fields, with special focus on the  RG flow interpretation of such a quantity. In Appendix \ref{app:sphericalHarmonics}, we provide a summary of useful properties of  spinor and spinor-vector spherical harmonics; in Appendix \ref{app:lattice}, we review the real-time numerical method necessary for the simulations and provide an analytic proof of the divergence of the sum over angular momentum modes when computing the EE with a radial discretization for fermionic fields in $d=4$; in Appendix \ref{app:subsec:ScalevsConf}, we  show that the correlators of gauge invariant operators display conformal symmetry, by identifying the field strength as a conformal primary and explicitly computing its two-point function. In addition, we provide a review of the embedding space  formalism, and give the explicit expression for the six-dimensional terms that produce the two-point function of a gauge invariant spin 3/2 conformal primary field in four dimensions.

\section{Entanglement entropy of a Dirac field in a sphere}
\label{sec:Dirac}
In this section, we revisit the entanglement entropy of a massless Dirac field in a sphere. We start by decomposing the Dirac Hamiltonian in $d=4$ in a spherical wave basis composed of spinor spherical harmonics and by integrating over the angular variables. This procedure dimensionally reduces the problem to a tower of $d=2$ fermions with a radial potential quadratic in the fields whose entanglement entropy can be computed numerically on a finite line.  We discover that the area term in the entanglement entropy  obtained by summing over the entropy of each mode diverges for the radial regularization. Notwithstanding, using the regularized entanglement entropy defined via the mutual information, the calculated logarithmic and area terms are in accordance with the literature.
\subsection{Decomposition in spherical harmonics}
\label{subsec:Diracmodes}
The massless Dirac Hamiltonian in $(3+1)$ spacetime dimensions is given by\footnote{We use Lorentz spacetime  indices $\mu , \, \nu,\dots=0,\dots,3$ and signature $\eta_{\mu \nu}=(+,-,-,-)$, Lorentz spatial indices $a,b,\dots=1,2,3$ and spinorial indices $i,j,\dots=1,\dots,4$. In addition, time coordinate  $t$ dependence is omitted as we are interested in computing entanglement entropies over constant time slices. }
\begin{equation}
H=\int d^3x\,\,\psi^{\dagger}(\vec{x})(\vec{\alpha}\cdot \vec{p}) \psi(\vec{x}),
\label{eq:diracham}
\end{equation}
where $(\vec{\alpha})^a=\gamma^0\gamma^a$ is defined in terms of gamma matrices   and $\vec{p}=-i\vec{\nabla}$ is the momentum operator. As is well known, choosing the Weyl representation of the gamma matrices
\begin{equation}
\gamma^\mu=\begin{pmatrix}
 0 & \sigma_R^\mu\\ 
\sigma_L^\mu & 0 
\end{pmatrix} \,,\quad{\sigma}_L^{\mu}\equiv(\mathbb{I}_2,-\sigma^i)\,,\quad {\sigma}_R^{\mu}\equiv(\mathbb{I}_2,\sigma^i)\:, \label{defweyl}
\end{equation}
decouples the Hamiltonian (\ref{eq:diracham}) as
\begin{equation}
H=H_L + H_R,\quad H_{\frac{L}{R}} = \mp  \int d^3x\,\,\psi_{\frac{L}{R}}^{\dagger}(\vec{x})(\vec{\sigma}\cdot \vec{p}) \psi_{\frac{L}{R}}(\vec{x}),
\label{eq:HamChiral}
\end{equation}
with $\psi_{L}$ and  $\psi_{R}$ being two-components Weyl spinors that represent left and right-handed fermions, respectively. These can be obtained from the original  spinor $\psi (\vec{x})$ by acting with with chirality projection operators defined in terms of $\gamma^5=i\gamma^0\gamma^1\gamma^2\gamma^3 $ as
\begin{equation}
\frac{1- \gamma^5}{2}\psi(\vec{x})=\begin{pmatrix}
\psi_L(\vec{x})\\ 
0\end{pmatrix} \,, \quad \frac{1+ \gamma^5}{2}\psi(\vec{x})=\begin{pmatrix}
0\\ 
\psi_R(\vec{x})\end{pmatrix}. \label{eq:chiral}
\end{equation}

The canonical phase space of the theory is spanned by the Weyl spinors  $\psi_{L}$ and  $\psi_{R}$ as canonical coordinates together with their corresponding conjugates $\psi^\dagger_{L}$ and  $\psi^\dagger_{R}$ as canonical momenta or viceversa. The corresponding canonical anticommutation relations at equal times are given by 
\begin{equation}
\big\{\big(\psi_{\frac{L}{R}}(\vec{x})\big)_i,\big(\psi^{\dagger}_{\frac{L}{R}}(\vec{x}\,')\big)_j\big\}= \delta_{ij}\delta^{(3)}(\vec{x}-\vec{x}\,'),
\label{eq:anticompsi}
\end{equation}
where the spinorial indices run in the subset $i,j=1,2$ for the Weyl spinors.
\bigskip

 Taking into account the rotational symmetry of the problem at hand, it is useful to perform a dimensional reduction of the chiral Dirac Hamiltonian over the sphere. For simplicity, we perform this over the left-handed Hamiltonian, but such procedure can be likewise applied for the right-handed one. To begin with, we expand the spinor $\psi_L$ in the spherical wave basis given by 
\begin{equation}
\psi_L (\vec{x}) =\sum_{\kappa \: \mu}\left(\frac{\p_{\kappa \mu}(r)}{r}\right)\Omega_{\kappa  \mu }(\theta,\varphi)\:,\:\:\:\:\:\:\psi^{\dagger}_L (\vec{x})=\sum_{\kappa \: \mu} \left(\frac{\p^{*}_{\kappa \mu}(r)}{r}\right)\Omega^{\dagger}_{\kappa  \mu }(\theta,\varphi) \:,
\label{eq:ModexpanDir}
\end{equation}
where $\Omega_{\kappa \mu}(\theta,\varphi)$ are the  spinor spherical harmonics 
 defined in  \cite{Szmytkowski:2007mzc} via the scalar spherical harmonics $Y_{lm}$ as
\begin{equation}
\Omega_{\kappa \mu}(\theta,\varphi)= \begin{pmatrix}
 \text{sgn}(-\kappa)\sqrt{\frac{\kappa+\frac{1}{2}-\mu}{2\kappa+1}} \: Y_{l,\mu - 1/2}(\theta,\varphi) \\ 
\sqrt{\frac{\kappa+\frac{1}{2}+\mu}{2\kappa+1}} \: Y_{l,\mu + 1/2}(\theta,\varphi) 
\end{pmatrix}\,. \label{eq:spinorhar}
\end{equation}
The label $\kappa$ is related to the orbital angular momentum $l$ by $\kappa=|l+\frac{1}{2}|-\frac{1}{2}$ and the total angular momentum $j=|\kappa|-\frac{1}{2}$ taking the discretized values $\kappa=\pm 1, \pm 2, \ldots $ with projections  $\mu = -|\kappa|+\frac{1}{2},-|\kappa|+\frac{3}{2},\ldots,|\kappa|-\frac{1}{2}$. For a brief review of useful properties of spinor spherical harmonics, see Appendix \ref{app:spinorh}.

\bigskip
\noindent Replacing  (\ref{eq:ModexpanDir}) into the left-handed Hamiltonian (\ref{eq:HamChiral}) gives
\begin{equation}
H_L=-i\sum_{\kappa '  \mu' \kappa \mu}\int d^3x\, \left[\left(\frac{\p^{*}_{\kappa' \mu'}}{r}\right)\Omega^{\dagger}_{\kappa'  \mu' }\right] \left(\partial_r+\frac{\kappa+1}{r} \right)\left[\left(\frac{\p_{\kappa \mu}}{r}\right)\Omega_{-\kappa  \mu }\right].
\label{HLD}
\end{equation}
Further, the orthonormality of the spinor spherical harmonics enables us to integrate out the angular dependence. Thus, the $(3+1)$ dimensional Hamiltonian is reduced to an effective tower of $(1+1)$ dimensional Hamiltonians in the half-line labeled by $(\kappa,\mu)$. The corresponding symmetrized real version of (\ref{HLD}) is 
\be
H_L^{\text{sym}}=-\frac{i}{2} \sum_{\kappa\mu}\int_0^{\infty} dr\,\, \left[\p^*_{-\kappa \mu}  \p'_{\kappa\mu}-\p^{*}{} '_{\kappa\mu}\p_{-\kappa\mu}+ \frac{\kappa}{r} \left(\p^*_{-\kappa \mu}\p_{\kappa\mu} - \p^*_{\kappa \mu}\p_{-\kappa\mu}\right) \right]\:,
\label{eq:dirac_ham_sym}
\ee
where we have used the notation $'$ to denote radial derivatives acting over radial functions (for example, $P'_{\kappa \mu}(r)= \partial_r P_{\kappa \mu} (r)$). We will use this notation throughout the paper.

The statistics for $(1+1)$ dimensional complex fields $\p_{\kappa \mu}$ can be recovered using the projections of the original spinor fields over the spinor spherical harmonics as  
\bea
\p_{\kappa \mu}(r)=r\int_{S^2} d\Omega \: \psi_L(\vec{x})\Omega^{\dagger}_{\kappa \mu}(\theta,\varphi),\quad\p_{\kappa \mu}^{*}(r)=r\int_{S^2} d\Omega \: \psi_L^{\dagger}(\vec{x})\Omega_{\kappa \mu}(\theta,\varphi).
\eea
Then, from (\ref{eq:anticompsi}), we get the equal time anticommutation relations 
\begin{equation}
\{(\p_{\kappa ' \mu '}(r'))_i,(\p_{\kappa \mu}^{\dagger}(r))_j\}= \delta_{ij}\delta_{\kappa \kappa'}\delta_{\mu \mu'}\delta(r-r').
\label{eq:commutF}
\end{equation}
However, we can reinterpret the $(1+1)$ dimensional fields  $\p_{\kappa \mu}$. To understand how $\p_{\kappa \mu}$ couples with $\p_{-\kappa \mu}$, one can rewrite $H_L^{\text{sym}}$ as a sum over positive modes $\kappa>0$. This leads to a theory described by a $(1+1)$ dimensional two-component spinor field $\Psi_{\kappa \mu}$ whose Hamiltonian is
\bea 
H_L^{\text{sym}}=\sum_{\kappa=1}^{\infty} \sum_{\mu}\int_0^{\infty} dr\:\Psi^{\dagger}_{\kappa \mu}(r)\left(\alpha_r\:\overset{\leftrightarrow}{p}_{r}+\frac{\kappa}{r}\beta_r\right)\Psi_{\kappa \mu}(r),  \quad \Psi_{\kappa \mu}(r)=\begin{pmatrix}
\p_{-\kappa \mu}(r) \\ \p_{\kappa \mu} (r)
\end{pmatrix}.   
\label{eq:Ham_compact_notation}
\eea
The induced radial representation gives $\alpha_r=\sigma^1$ and  $\beta_r=\sigma^2$, where $\overset{\leftrightarrow}{p}_r=-\frac{i}{2}(\vec{\partial}_r-\cev{\partial}_r )$ is the symmetrized momentum operator on the radial direction. The sum over $\kappa$ can be thought of as a sum over all modes with definite angular momenta $j=\frac{1}{2},\frac{3}{2},\dots$, recalling that  $j=|\kappa|-\frac{1}{2}$. Also, from (\ref{eq:commutF}) the spinor field $\Psi_{\kappa \mu}$ obeys the equal time canonical anticommutation relations 
\bea
\left\{\big(\Psi_{\kappa' \mu '}(r')\big)_i,\big(\Psi^{\dagger}_{\kappa \mu}(r)\big)_j\right\}=\delta_{ij} \delta_{\kappa \kappa'}\delta_{\mu \mu'}\delta(r-r').
\label{eq:CommutDirac}
\eea
To sum up, we have obtained a tower of Dirac fields in the half-line with a position-dependent quadratic term $\frac{\kappa}{r}$ that appear as a consequence of the Kaluza-Klein reduction on the transverse sphere $S^2$.

The Hamiltonian (\ref{eq:Ham_compact_notation}) together with (\ref{eq:CommutDirac})  represent the main results of this subsection. They will prove useful when performing a numerical computation of the entanglement entropy in Section \ref{subsec:latticeDirac}, as well as when comparing with the analogue mode decomposition for the Rarita-Schwinger field in Section \ref{subsec:logRarita}.  The effective radial equations for the Dirac field have been obtained via different methods and for arbitrary dimensions in the literature, for instance by \cite{Lopez-Ortega:2009flo}\footnote{\noindent The matching with (\ref{eq:Ham_compact_notation}) follows first from a unitary  map of the spinor field and a subsequent axial rotation.  The  transformed field satisfies the equations of motion like those in  \cite{Lopez-Ortega:2009flo}.} Further study regarding the modular Hamiltonians of these (1+1) fermionic modes and their contribution to the universal coefficients of the EE in general dimensions, as performed for scalars in \cite{Huerta:2022tpq}, will appear in \cite{MarinaGuido}.

\subsection{Entanglement entropy and logarithmic coefficient}
\label{subsec:latticeDirac}
This subsection is devoted to the study of the EE  associated to a spherical region of radius $R$ for a free massless Dirac fermion in $d=4$ space-time dimensions. We will follow a similar procedure as for the scalar field in \cite{Srednicki:1993im}. Therefore, we are particularly interested in the EE associated with the tower of two-dimensional Dirac fermions with a radial potential in the half-line. Given that the Hamiltonian (\ref{eq:Ham_compact_notation}) is expressed as a sum over different modes $(\kappa,\mu)$, the resulting expression for the EE $S(R)$ of Dirac fermions on the sphere holds after summing over each of them 
\begin{equation}
    S(R)=S_L(R)+S_R(R)=2\sum_{\kappa=1}^{\infty}\sum_{\mu}S_{\kappa}(R)=2\sum_{\kappa=1}^{\infty}(2\kappa)S_{\kappa}(R).
    \label{eq:sumDiracAngular}
\end{equation}
The first factor of two is due to the left/right chiralites in (\ref{eq:HamChiral}), and the factor $2\kappa$ accounts for the multiplicity of $\mu$-projections for a given $\kappa$, because the EE $S_{\kappa}(R)$  associated to each of the modes in (\ref{eq:Ham_compact_notation}) is $\mu$-independent\footnote{The fact that the contributions $S_{\kappa}(R)$  are independent of $\mu$ follows from the expression (\ref{eq:Ham_compact_notation}), given that for a given $\kappa$ all the modes have the same Hamiltonian. In other words, for a fixed $\kappa$, all the the modes have the same dynamics, as well the same algebra (\ref{eq:CommutDirac}), yielding the same entanglement entropy.}.  Note that the sum (\ref{eq:sumDiracAngular}) is a formal expression, since this bare series may not be convergent  for some choices of regularization. We will argue in what follows that this  is  the case for the radial discretization introduced for a scalar field in \cite{Srednicki:1993im}, when applied to a fermion field.

 For quadratic Hamiltonians such as (\ref{eq:Ham_compact_notation}), the vacuum is a Gaussian state. This implies that Wick's theorem holds, and all non-zero multipoint correlators can be obtained from the two-point function. In the light of this, the EE $S_\kappa (R)$ can be computed numerically by the real-time method introduced in \cite{2003Peschel,Casini:2009sr}.  The idea is to consider the Hamiltonian on the lattice, with a total of $M$ lattice sites, as well as the correlator $C_{mn}=\langle \psi^{\dagger}_m\psi_n \rangle$,
 where $\psi_n$ stands for the lattice discretization of $\Psi_{\kappa \mu}(r)$. Hence, considering a spherical region of radius $R=N+\frac{1}{2}$, the EE $S_{\kappa}(N,M)$  depends exclusively on the eigenvalues of $C_{mn}$ when restricted to $m,n=1,\dots,N$. A holistic explanation of this standard procedure is found in Appendix \ref{app:numerical}. 
 
\begin{figure}[t]
    \centering
    \includegraphics[width=\textwidth]{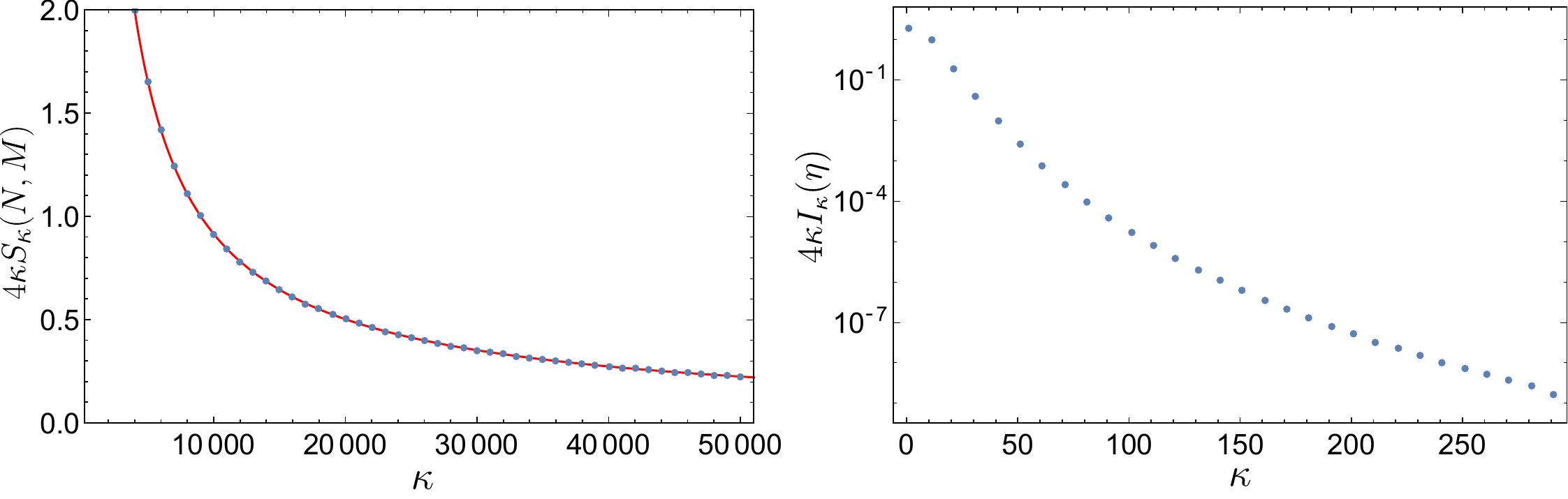}
    \caption{Left plot: Entanglement entropy for a massless Dirac field in $d=4$ as a function of angular momentum $\kappa$ for large $\kappa$. The dotted points indicate the lattice simulation data and the red continuous line indicates the fit function $f(\kappa)=a\frac{\log(\kappa)}{\kappa}+b\frac{1}{\kappa}$. The fitted values $a=1274.94$ and $b=-2595.83$ match with relative error of $0.005\%$ and $0.03\%$, respectively, with the coefficients of  (\ref{eq:tailDirac}). The numerical simulation was done with $M=100$ physical sites and considered a spherical region of fixed radius $R=N+\frac{1}{2}=50.5\:$. Right plot: Mutual information in logarithmic scale, for a massless Dirac field in $d=4$, as a function of angular momentum $\kappa$ for fixed $\eta$. In stark contrast with the entanglement entropy (in radial regularization), the mutual information is exponentially suppressed for large $\kappa$. The simulation was done for fixed $\eta=9.5$ and $M=100$ physical sites.}
    \label{fig:tailDirac}
\end{figure}

 The contributions from modes with  large angular momentum $\kappa$  are relevant for the convergence of the sum (\ref{eq:sumDiracAngular}). The entropies $S_\kappa(N,M)$ are independent of the infrared cutoff $M$ for $\kappa \gg M$ and behave as
\begin{equation}
4\,\kappa \,S_{\kappa}(N,M)\sim 8\xi(N)\,\left(\frac{\log(\kappa)}{\kappa}\right)+4\xi(N)\big[1-\log(\xi(N))\big]\,\left(\frac{1}{\kappa}\right)\,+\ldots,
\label{eq:tailDirac}
\end{equation}
where $\xi(N)$ is given by
\begin{equation}
\xi(N)=\frac{N^2(N+1)^2}{4(2N+1)^2}.
\end{equation}
For an analytic derivation of this asymptotic behavior, see Appendix \ref{app:Largek}, and for a numerical verification, refer to the left plot of Fig. \ref{fig:tailDirac}.  As a consequence of (\ref{eq:tailDirac}), the sum introduced in (\ref{eq:sumDiracAngular}) diverges for an arbitrary cutoff $M$. This behavior is in stark contrast to what happens with a scalar field in $d=4$, where the radial discretization of the EE is convergent. This is so because the leading term of the entropy of each mode decays as  $\sim\frac{\log(l)}{l^3}$ for large angular momentum $l$ \cite{Srednicki:1993im}. The physical interpretation of this phenomena is that for a Dirac field, the radial discretization is not sufficient for inducing a finite area coefficient $c_2$  in (\ref{eq:EntropyCFT}). This situation is not unfamiliar, as a similar scenario occurs for scalar fields in $d>4$ \cite{Riera:2006vj}. 
 
\begin{figure}[t]
    \centering
    \includegraphics[width=0.6\textwidth]{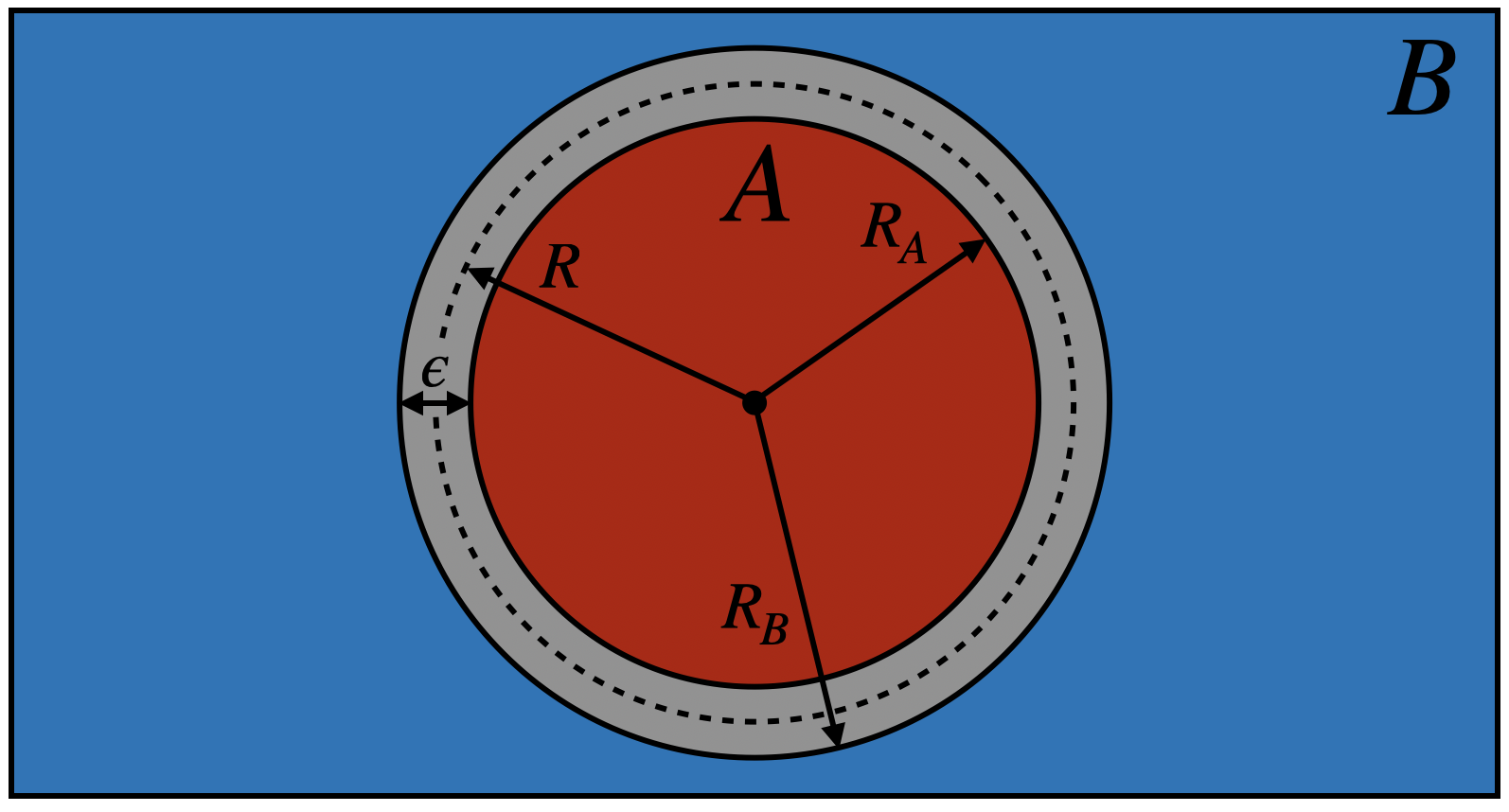}
    \caption{Set-up used to compute the mutual information $I(A,B)$. The region $A $ is a sphere of  radius $R_A$ and $B$ is the complementary region of a sphere of radius $R_B$ that contains $A$. Both radii can also be parameterized in terms of new variables $\epsilon$ and $R$ as $R_A=R-\epsilon/2$ and $R_B=R+\epsilon/2$.}
    \label{fig:Mutualfig}
\end{figure}

\begin{figure}[t]
    \centering
    \includegraphics[width=0.99\textwidth]{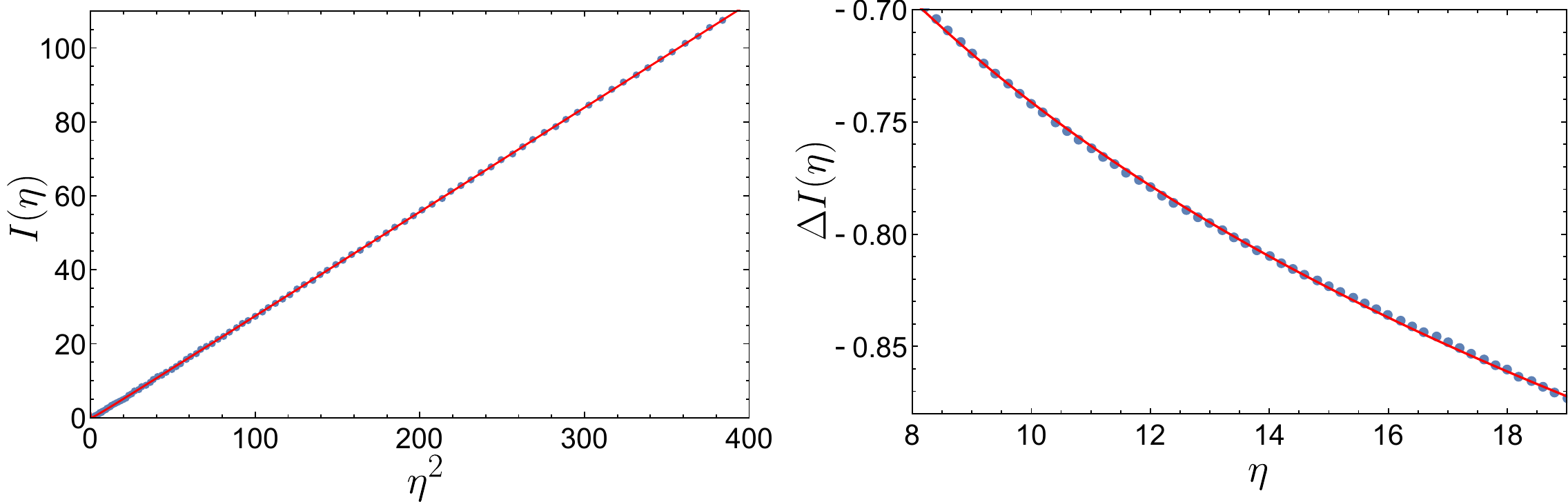}
    \caption{Left plot: Mutual information $I(\eta)$ for a Dirac field as a function of $\eta^2$ for several values of $R_1$, $R_2$, with $M=200$ physical sites. The dotted points indicate the lattice simulation data and the continuous red line indicates the fitting function $f(\eta)=2\,c_2^{\text{fit}}\eta^2+2\,c^{\text{fit}}_{\text{log}}\log(\eta)+ c_0^{\text{fit}}$. Right plot: Mutual information with the area term subtracted, $\Delta I(\eta)=I(\eta)-2c_2^{\text{fit}}\eta^2$,  as a function of $\eta$. The fit exhibits a subleading logarithmic contribution that goes like $\Delta I(\eta)=2\,c^{\text{fit}}_{\text{log}}\log(\eta)+ c_0^{\text{fit}}$.}
    \label{fig:MergeMutual}
\end{figure}

The area coefficient $c_2$ is regularization dependent, so this issue can be overcome by choosing another regulator. Conveniently, the mutual information gives a geometrical prescription for computing a regularized EE \cite{Casini:2015woa,Casini:2015dsg}. This construction relies on the mutual information of a spherical region $A$ of radius $R_A$ and a region $B$ defined as the complement of a sphere of radius $R_B$, with $R_B>R_A$, that contains the region $A$  (see Fig. \ref{fig:Mutualfig}). The mutual information is defined as usual
\begin{equation}
    I(A,B)=S(A)+S(B)-S(A\,\cup\,B).\label{midef}
\end{equation}
Purity of the vacuum state implies that the entropy of a region coincides with the entropy of its complement. This means that $S(B)$ can be computed as the entropy of a sphere of radius $R_B$ and $S(A\,\cup\,B)$ as the entropy of an annular region defined between $R_A$ and $R_B$. We can introduce a geometric regularization by expressing the radii in terms of two new variables $R$ and $\epsilon$ such that $R_A=R-{\epsilon}/{2}$ and $R_B=R+{\epsilon}/{2}$. This motivates the following definition
\begin{equation}
    \eta\equiv \frac{R}{\epsilon}=\frac{R_A+R_B}{2(R_B-R_A)}.
\end{equation}
In this vein, for $\eta \gg 1$ the mutual information gives a prescription for the regularized EE in the form of
\begin{equation}
    S_{\text{reg}}(\eta)=\frac{I(\eta)}{2}=s_2 \eta^2+c_{\text{log}}\log(\eta)\,,\quad \eta \gg 1\,. \label{eq:reg-ent}
\end{equation}
In Fig.\ref{fig:MergeMutual}, we show numerical simulations in the range $8<\eta<20$. In opposition to the slowly decaying behavior of the EE as a function of $\kappa$, we observed an exponentially suppressed decay for large $\kappa$ for the mutual information\footnote{This same behavior was also numerically observed for the Maxwell field in \cite{Casini:2015dsg}.} (see right plot of Fig. \ref{fig:tailDirac}). This is because the mutual information is less sensitive to ultraviolet
contributions than the EE. For the maximum lattice size considered $M_{\text{max}}=200$, we noted that it was sufficient to sum up to $\kappa_{\text{max}}=250$ in (\ref{eq:sumDiracAngular}). The extraction of a numerical value for $s_2$ and $c_{\text{log}}$ requires the elimination of the dependence on the infrared cutoff $M$. Therefore, by fitting (\ref{eq:reg-ent}) for lattices with size $M=100, 125, 150, 175, 200$, the fitted coefficients were extrapolated to the continuum by using the functions 
\be s_2(M) = s_2^{\infty}+\frac{a}{M^2}\,,\quad c_\text{log}(M)=c_\text{log}^{\infty}+\frac{b}{M^2}\,.\ee
The resulting $s_2^{\infty}$ and $c_\text{log}^{\infty}$, interpreted as the continuum limit results, are
\be s^{\infty}_2=0.14145. \,, \quad  c_{\text{log}}^{\infty}=-0.12285\,. \ee
The logarithmic coefficient matches with the expected value  $c_{\text{log}}(1/2)=-\frac{11}{90}=-0.12222 \dots$ with 0.5$\%$ of relative error. The area term also coincides with an approximation present in the literature\footnote{ The area coefficient of the regularized EE can be obtained as 
 $s_2=4\pi(2\kappa_4)=0.13521\dots$ where $\kappa_4=0.00538\dots$ represents the area coefficient of the EE associated to a region between two parallel plates \cite{Casini:2015dsg,Casini:2009sr}. } with 4.6$\%$ of relative error.  
\section{Free Rarita-Schwinger theory}
\label{sec:RaritaSTheory}


The objective of this section is to study the  Rarita-Schwinger field \cite{Rarita:1941mf,freedman2012,Weinberg:2000cr} as a free quantum field theory describing massless spin $3/2$ degrees of freedom in $d=4$ spacetime dimensions.  The free massless field theory is well defined at the classical and quantum level (see \cite{freedman2012} and references therein). In this vein, we start by reviewing the canonical quantization considering a vector-spinor gauge field\footnote{The vector-spinor gauge field does not transform under an irreducible representation of the four-dimensional Lorentz group, and a priori describes propagating spin $3/2$  degrees of freedom together with spin $1/2$ ones (see for instance \cite{freedman2012,Weinberg:2000cr} and \cite{Valenzuela:2022gbk,Valenzuela:2023aoa} for a more recent discussion).}. Then, we show how such procedure is implemented over the gauge invariant phase space spanned by spinor two-form field strength defined from the original gauge field. This is known to  only include the spin $3/2$ degrees of freedom \cite{Allcock:1976qp,Deser:1977ur}. 

\subsection{Canonical formulation}
\label{subsec:Raritacanonical}
A theory with  propagating spin $3/2$ fermions can be described by a vector-valued spinor field $\psi_\nu (x)$ with $\nu=0,1,2,3$ in $d=4$. The dynamics of such fields  can be given by the Rarita-Schwinger Lagrangian \cite{freedman2012,Weinberg:2000cr}. The latter, in the strict massless limit, is written as \cite{Adler:2015yha,Adler:2017lki}
\begin{equation}
\mathcal{L}=\frac{i}{2}\epsilon_{ \nu \rho \sigma \lambda}\bar{\psi}^{\nu}\gamma^5\gamma^{\rho}\partial^{\sigma}\psi^{\lambda}\:,
\label{eq:RaritaLagTot}
\end{equation}
where $\epsilon_{ \nu \rho \sigma \lambda}$ is the $4$-dimensional Levi-Civita tensor, $\gamma^{\rho}$ and $\gamma^{5}=i\gamma^0\gamma^1\gamma^2\gamma^3$ are the  usual gamma matrices and $\bar{\psi}_{\mu}=i\psi_{\mu}^{\dagger}\gamma^0$. 
In analogy with spin $1/2$ fermions, we can choose the Weyl representation such that the Lagrangian decouples into two independent sectors
\begin{equation}
\mathcal{L}=\mathcal{L}_{L}+\mathcal{L}_{R}\,, \quad \mathcal{L}_{\frac{L}{R}}=\frac{1}{2}\epsilon_{abc}\Big(\psi_{\frac{L}{R}}^{a\dagger}\sigma^b \dot{\psi}_{\frac{L}{R}}^c\pm\psi_{\frac{L}{R}}^{a\dagger}\partial^b \psi_{\frac{L}{R}}^c+\psi^{0\dagger}_{\frac{L}{R}}\sigma^a\partial^b\psi^c_{\frac{L}{R}}-\psi^{a\dagger}_{\frac{L}{R}}\sigma^b\partial^c \psi^{0}_{\frac{L}{R}}\Big),
\label{rslr}
\end{equation}
where  the Weyl spinors  $\psi^\mu_{L}$ and $\psi^\mu_{R}$ are defined using the chiral projectors as in (\ref{eq:chiral}), and $\epsilon_{abc}$ represents the 3-dimensional Levi-Civita tensor.

From here, we will only deal with the theory of left-handed fermions. However, the following procedure can also be straightforwardly generalized to right-handed fermions too. First,
the  timelike component $\psi^0_L$ of the field only appears in the  last two terms of (\ref{rslr}) as a Lagrange multiplier. On the other hand, the space-like components $\psi^a_L$ of the field appear as dynamical variables associated with the canonical momenta
\begin{equation}
\pi_a^{L}=\frac{\partial \mathcal{L}}{\partial\dot{\psi^a_{L}}}=\frac{1}{2}\epsilon_{abc}\psi^{b\dagger}_{L}\sigma^c.
\end{equation}
At a quantum level, these anticommute with the canonical coordinates at equal times, giving
\be 
\{(\psi^{a}_L(\vec{x}))_i,(\pi^{b}_L(\vec{x}'))_j\}=-i\delta^{ab}\delta_{ij}\,\delta^{(3)}(\vec{x}-\vec{x}').
\ee
Conveniently, we can use  $\psi_L^{a\,\dagger }= \,i \pi^a_L - \varepsilon^{abc}\pi^{L}_b \sigma_c\,,$
to obtain a more useful expression in terms of the conjugated field 
\be 
\{(\psi^{a}_L(\vec{x}))_i,(\psi^{b\,\dagger  }_L(\vec{x}'))_j\}=(\sigma^b\sigma^a)_{ij}\delta^{(3)}(\vec{x}-\vec{x}').
\label{eq:canonantC}
\ee
In this setup, we can compute the Hamiltonian, via a Legendre transformation, to be
\begin{equation}
\mathcal{H}_{L}=\pi^{L}_a\dot{\psi}^a_{L}-\mathcal{L}_L=- \frac{1}{2}\epsilon_{abc}\Big(\psi_{L}^{a\dagger}\partial^b \psi_{L}^c +\psi^{0\dagger}_{L}\sigma^a\partial^b\psi^c_{L}-\psi^{a\dagger}_{L}\sigma^b\partial^c \psi^{0}_{L}\Big).
\label{rsham}
\end{equation}
Then, we are  allowed to derive the constraints from the equations of motion of the timelike component of the fields. These yield the Gauss-law-like constraints
\begin{equation}
2 \,\partial_a\pi_L^a= \varepsilon_{abc} \sigma^a \partial^b \psi^{c\:\dagger}_{L}=0 \,,\quad 2 \,\partial_a\pi^{a\,\dagger}_L= \varepsilon_{abc} \sigma^a \partial^b \psi^{c}_{L}=0 \,.
\label{eq:constraints}
\end{equation}
In the massless theory, there is a gauge symmetry that can be described in terms of an arbitrary local spinor function  $\chi(x)$ as $\psi^{\nu} (x) \rightarrow \psi^{\nu}(x) + \partial^{\nu}\chi(x)$. Therefore, acting over the left-handed spinor as the chiral projection of this symmetry
\begin{equation}
    \psi_L^{\nu} (x) \rightarrow \psi_L^{\nu}(x) + \partial^{\nu} \chi_L (x)\:.
    \label{gaugers}
\end{equation}
This implies that neither the dynamical fermion field $\psi_L^{a} (x)$ nor any of its bilinears can be considered observables. We will deal with this issue in the following section.
\subsection{Gauge invariant phase space}
\label{subsec:GaugeInvaCanformulatiomm}
 The formulation of the Rarita-Schwinger theory in terms of phase space gauge invariant variables, makes the computation of the EE easier to understand, as it can be formulated in terms of algebras of local observables attached to regions consisting solely of spin 3/2 degrees of freedom. To begin, we consider the gauge-invariant operators spanned from the spinor field strength associated with the vector-spinor field \cite{Allcock:1976qp,Deser:1977ur,freedman2012}. Specifically, this is a spinor 2-form given by
\begin{equation}
    F_{\mu \nu}=\partial_{\mu}\psi_{\nu}-\partial_{\nu}\psi_{\mu}. 
    \label{eq:fieldStrenght}
\end{equation}
The dynamics obtained from (\ref{eq:RaritaLagTot}), can be recovered in  terms of the spinor field strength as
\begin{equation}
\gamma^{\mu}F_{\mu \nu}=0,\:\:\:\:\:\:\:\:\:\:\:\: \partial^{\mu}F_{\mu \nu}=0\:.
\label{eq:EOMF}
\end{equation}
These conditions secure that field strength $F$ belongs to the representation $(3/2,0)\oplus (0,3/2)$ of the Lorentz group \cite{Allcock:1976qp}. Moreover, the irreducibility condition $\gamma^{\mu}\gamma^{\nu}F_{\mu \nu}=0 $, follows  trivially from (\ref{eq:EOMF}). Physically the irreducibility condition eliminates the spin $1/2$ degrees of freedom that transform in the representation $(1/2,0)\oplus (0,1/2)$ \cite{Weinberg:2012mz}. Therefore, $F$ is a gauge invariant operator that describes the propagation of free massless particles of spin $3/2$\footnote{It would be interesting to understand how this description overlaps with a recent discussion about the propagating spin-$3/2$ and spin-$1/2$ degrees of freedom \cite{Valenzuela:2022gbk,Valenzuela:2023aoa} in terms of local algebras of observables.}. \\

 Although  the spinor field strength $F$ is a generator of the gauge invariant phase space, we can also use the Levi-Civita tensor to produce new tensors. That is, one should also include the dual spinor 2-form given by 
\begin{equation}
    F^*_{\mu \nu}=\frac{1}{2}\epsilon_{\mu \nu \sigma \rho}F^{\sigma \rho}\:.
\end{equation}
However, it is clear that both spinor 2-form fields $F$ and $ F^*$ span the same phase space. More precisely, considering the equations of motion $\gamma^{\mu}F_{\mu \nu}=0$, we can explicitly check that 
\begin{equation}
    F_{\mu \nu}+i\gamma^5 F^*_{\mu \nu}=0\:.
    \label{eq:dualitFFdual}
\end{equation}
Indeed, the description in terms of $F$ and $ F^*$  are interchangeable since both obey the same equations of motion as well as divergenceless conditions. Namely,
\begin{equation}
\gamma^{\mu}F^*_{\mu \nu}=0,\:\:\:\:\:\:\:\:\:\:\:\: \partial^{\mu}F^*_{\mu \nu}=0\:.
\end{equation}
The irreducible representations $(3/2,0)$ and $(0,3/2)$ are described by the self-dual and anti-self dual parts of the field strength, respectively. These are defined to have  eigenvalues $\pm i$ with respect to the Hodge star operator $*$ in Minkoswki spacetime
\begin{equation}
    *(F_{\mu \nu}\pm i F^*_{\mu \nu})=(\mp i) (F_{\mu \nu}\pm i F^*_{\mu \nu})\:. 
\end{equation}
In this way, the equation (\ref{eq:dualitFFdual}) becomes clear as it dictates that the  self/anti-self dual forms are related to the left and right-handed parts of the vector-spinor field strength
\begin{equation}
  \frac{1}{2}(F_{\mu \nu}+i F^*_{\mu \nu})=\left(\frac{1-\gamma^5}{2}\right)F_{\mu \nu} = \begin{pmatrix}F^L_{\mu \nu}(\vec{x})\:,\\ 
0\end{pmatrix}   ,\quad  F^L_{\mu \nu}= \partial_{\mu}\psi^L_{\nu}-\partial_{\nu}\psi^L_{\mu}\:,
\end{equation}
\begin{equation}
  \frac{1}{2}(F_{\mu \nu}-i F^*_{\mu \nu})=\left(\frac{1+\gamma^5}{2}\right)F_{\mu \nu} = \begin{pmatrix}0\\ 
F^R_{\mu \nu}(\vec{x})\end{pmatrix}   ,\quad  F^R_{\mu \nu}= \partial_{\mu}\psi^R_{\nu}-\partial_{\nu}\psi^R_{\mu}\:.\label{FR}
\end{equation}
 Also, in the Weyl representation of the gamma matrices (\ref{defweyl}), the equations of motion and divergenceless conditions are
\begin{equation}
{\sigma}_L^{\mu}F^L_{\mu \nu}=0,\:\:\:\:\:\:\:\:\:\:\:\: \partial^{\mu}F^L_{\mu \nu}=0\:,\:\:\:\:\:\:\:\:\:\:\:\:\sigma^{\mu}_RF^R_{\mu \nu}=0,\:\:\:\:\:\:\:\:\:\:\:\: \partial^{\mu}F^R_{\mu \nu}=0\:. \label{rfl}
\end{equation}
 In addition, from the canonical anticommutation relations (\ref{eq:canonantC}) we can ascertain that the components of $F_{ab}^{L}$ and $F_{ab}^{L\,\,\dagger}$ do  not anticommute
\begin{equation}
\{(F_{ab}^{L} (\vec{x}))^i,(F_{cd}^{L\,\,\dagger}(\vec{x}'))^j\}=[\sigma_g\sigma_f]^{ij} \,\delta_{ab}^{ef}\,\delta_{cd}^{gh} \,\partial_e\partial_g \,\delta^{(3)}(\vec{x}-\vec{x}'),\,\quad \delta_{ab}^{cd}=\delta_a^c \delta_b^d-\delta_a^d \delta_b^c\,. \label{esta}
\end{equation}
The remaining relations involving $F_{0a}^L$  follow from (\ref{rfl}).  The right-handed field strength (\ref{FR}) obeys analogous anticommutation relations. 

\section{Entanglement entropy of a Rarita-Schwinger field in a sphere}
\label{sec:RaritaS}
In this section, we study the EE of a free massless Rarita-Schwinger field in a sphere. Mimicking the procedure outlined in Section \ref{subsec:Diracmodes}, the Hamiltonian is decomposed in a spherical wave basis of vector-spinor spherical harmonics. By integrating over the angular variables, this course of action dimensionally reduces the problem to a Dirac Hamiltonian without the $j=\frac{1}{2}$ mode. This identification offers a method for evaluating the EE. Although the  decomposition is performed over the gauge-dependent vector-spinor field, we explicitly provide a gauge fixing condition that describes the field in terms of gauge invariant operators localized in the sphere. This condition also secures that the EE result corresponds solely to propagating spin $3/2$ degrees of freedom. We provide the logarithmic coefficient that appears in the EE and suggests a further generalization to higher free half-integer spin fields.

\subsection{Decomposition in spherical harmonics}
\label{subsec:Raritamodes}
To perform the mode decomposition for the  Rarita-Schwinger Hamiltonian, we need to introduce the notion of vector-spinor spherical harmonics. These can be defined by extending the  definition of the spinor spherical harmonics (\ref{eq:spinorhar}) using  vector spherical harmonics instead of scalar spherical harmonics. More specifically, we can write 
\begin{equation}
\vec{\Omega}^s_{\kappa \mu}(\theta,\varphi)= \begin{pmatrix}
 \text{sgn}(-\kappa)\sqrt{\frac{\kappa+\frac{1}{2}-\mu}{2\kappa+1}} \: \vec{Y}^s_{l,\mu - 1/2}(\theta,\varphi) \\ 
\sqrt{\frac{\kappa+\frac{1}{2}+\mu}{2\kappa+1}} \: \vec{Y}^s_{l,\mu + 1/2}(\theta,\varphi) 
\end{pmatrix} \,,\quad s=r,e,m
, \label{spi}
\end{equation}
where the index $s=r,m,e$ is not a Lorentz spatial index, instead it indicates the type of vector spherical harmonic, denoted as: radial $\vec{Y}^r_{lm}$, electric $\vec{Y}^e_{lm}$, and magnetic $\vec{Y}^m_{lm}$ vector spherical harmonics. These vector spherical harmonics are vectors themselves that can be used to expand vector-valued functions (see \cite{Casini:2015dsg} for an example in the context of entanglement entropy calculations). More specifically, they can be obtained from the usual spherical harmonics as
\begin{align}
\vec{Y}^r_{lm} (\theta,\varphi) & = {Y}_{lm} (\theta,\varphi)  \,\hat{r} \,,\,\, \:\:\:\: \:\:\:\:\:\:\:\:\:\:\:\:\:\:\:\:l\geq 0\:\:\:\: \, -l \leq m \leq l  \\
\vec{Y}^e_{lm} (\theta,\varphi) & = \frac{r \vec{\nabla}{Y}_{lm} (\theta,\varphi) }{\sqrt{l(l+1)}} \,,\,\,\:\:\:\: \:\:\:\:\:\:\:\:\:\:\:\:l\geq 1\:\:\:\: \, -l \leq m \leq l   \\
\vec{Y}^m_{lm} (\theta,\varphi) & = \frac{\vec{r}\times \vec{\nabla}{Y}_{lm} (\theta,\varphi) }{\sqrt{l(l+1)}} \,,\,\, \:\:\:\: \:\:\:\:\:\:l\geq 1\:\:\:\: \, -l \leq m \leq l  \:.
\end{align}
Note that  in (\ref{spi}) the $\kappa=-1$ mode corresponds to the $l=0$ mode, thus $\vec{\Omega}^e_{-1 \mu}$ and $\vec{\Omega}^m_{-1 \mu}$ are not well defined. Further properties of vector-spinor spherical harmonics, such as orthonormality, extend from the properties of vector spherical harmonics.  See Appendix \ref{app:vecspinorh} for a summary of useful expressions.

 To proceed, we expand the dynamical variables $\vec{\psi}_L$ using  vector-spinor spherical harmonics and the Lagrange multipliers $\psi_L^0$ using spinor spherical harmonics,
\begin{equation}
\vec{\psi_L} (\vec{x})  = \sum_{s\:\kappa \: \mu}\left(\frac{\p^s_{\kappa \mu}(r)}{r}\right)\vec{\Omega}^s_{\kappa  \mu }(\theta,\varphi) \,, \quad \psi^0_L (\vec{x})  =\sum_{\kappa \: \mu}\left(\frac{\p^0_{\kappa \mu}(r)}{r}\right)\Omega_{\kappa  \mu }(\theta,\varphi) \:.
\label{lmd1}
\end{equation}
We now replace the expansion (\ref{lmd1}) in the Rarita-Schwinger Hamiltonian coming from the density (\ref{rsham}), which in vector notation reads
\begin{equation}
H_L=-\frac{1}{2}\int d^3x\: \Big[\vec{\psi}^{\:\dagger}_{L}\cdot (\vec{\nabla}\times \vec{\psi}_{L})+\psi^{0\dagger}_{L}(\vec{\sigma} \cdot (\vec{\nabla}\times \vec{\psi}_{L}))-( (\vec{\nabla}\times \vec{\psi}^\dagger_{L}) \cdot \vec{\sigma} ) \psi^{0}_{L}\Big].
\label{rsvecham}
\end{equation}
 The resulting Hamiltonian for $\kappa=-1$ is trivially zero since the electric and magnetic modes are not defined. The  Hamiltonian recovered by summing over the remaining cases yields\footnote{We remind the reader that we use the notation $'$ to denote radial derivatives $\partial_r$ acting over functions that depend on the radial coordinate $r$.}
\begin{align}
 H_L =-&\frac{1}{2}\sum_{\kappa \neq -1 }\sum_\mu \int_0^{\infty} dr \:\left\{\left[\p_{\kappa  \mu }^{e*}{\p'}_{\kappa  \mu }^m -\p_{\kappa  \mu }^{m*}{\p'}_{\kappa  \mu }^e +\frac{\sqrt{\kappa(\kappa+1)}}{r} \left(\p_{\kappa  \mu }^{r*}\p_{\kappa  \mu }^m  + \p_{\kappa  \mu }^{m*}\p_{\kappa  \mu }^r \right)\right]+\right. \nonumber\\
 +& \sqrt{\frac{\kappa+1}{\kappa}} \p^{0*}_{\kappa  \mu } \left[\sqrt{\frac{\kappa-1}{\kappa+1}}\left({\p'}_{-\kappa  \mu }^m -  \frac{\kappa}{r} \p_{-\kappa  \mu }^m \right)-i \left( {\p'}_{\kappa  \mu }^e -\frac{ \sqrt{\kappa(\kappa+1)}}{r} \p_{\kappa \mu }^r  \right) \right]+ \label{rsvecham2}\\
+& \left. \sqrt{\frac{\kappa+1}{\kappa}} \left[\sqrt{\frac{\kappa-1}{\kappa+1}}\left({\p'}_{-\kappa  \mu }^{m*} -  \frac{\kappa}{r} \p_{-\kappa  \mu }^{m*} \right)+i  \left( {\p'}_{\kappa  \mu }^{e*} -\frac{ \sqrt{\kappa(\kappa+1)}}{r} \p_{\kappa \mu }^{r*}  \right) \right]  \p^{0}_{\kappa  \mu }\right\} \nonumber\:,
 \end{align}
 where we have considered the action of the differential operators in (\ref{rsvecham}) acting on $\vec{\psi}_{L}$ as
 \begin{equation}
     \vec{\nabla}\times \vec{\psi}_{L} =\frac{1}{r}\sum_{\kappa \: \mu} \left[  -\frac{\sqrt{\kappa(\kappa+1)}}{r} \p_{\kappa  \mu }^m  \vec{\Omega}^r_{\kappa  \mu }-{\p'}_{\kappa  \mu }^m \vec{\Omega}^e_{\kappa  \mu } +\left( {\p'}_{\kappa  \mu }^e -\frac{\sqrt{\kappa(\kappa+1)}}{r} \p_{\kappa  \mu }^r \right) \vec{\Omega}^m_{\kappa  \mu }\right] \:,
\end{equation}
 as well as the scalar product with the Pauli matrices
{\small \begin{equation} 
    \vec{\sigma}\cdot(\vec{\nabla}\times \vec{\psi}_{L}) = \frac{1}{r}\sum_{\kappa \: \mu} \sqrt{\frac{\kappa+1}{\kappa}}\left[  \left({\p'}_{\kappa  \mu }^m +  \frac{\kappa}{r} \p_{\kappa  \mu }^m \right)\Omega_{-\kappa  \mu }-i  \left( {\p'}_{\kappa  \mu }^e -\frac{ \sqrt{\kappa(\kappa+1)}}{r} \p_{\kappa \mu }^r  \right) \Omega_{\kappa  \mu } \right]   \:. 
 \end{equation}}
In this context, the constraints for each mode are enforced by the equations of motion of the $d=2$ dimensional multipliers $\p^{0}_{\kappa  \mu }$ and $\p^{0\:*}_{\kappa  \mu }$. From (\ref{rsvecham2}), we compute the constraint
\begin{equation}
{\p'}_{\kappa  \mu }^e -\frac{ \sqrt{\kappa(\kappa+1)}}{r} \p_{\kappa \mu }^r = - i\sqrt{\frac{\kappa-1}{\kappa+1}}\left({\p'}_{-\kappa  \mu }^m -  \frac{\kappa}{r} \p_{-\kappa  \mu }^m \right)\:.
\end{equation}
Replacing this in the Hamiltonian (\ref{rsvecham2}) we get a tower of $d=2$  dimensional Hamiltonians that only depend on the magnetic components of the original spinor
 \begin{equation}
H_L =\frac{i}{2}\sum_{\kappa \neq \pm 1,\,\mu } \int_0^{\infty} dr \: \sqrt{\frac{\kappa-1}{\kappa+1}}\left[{\p'}_{-\kappa  \mu }^{m *} {\p}_{\kappa  \mu }^m - \p_{\kappa  \mu }^{m*} {\p'}_{-\kappa  \mu }^m +   \frac{\kappa}{r} \left(  \p_{\kappa  \mu }^{m*}\p_{-\kappa  \mu }^m  -\p_{-\kappa  \mu }^{m*}{\p}_{\kappa  \mu }^m\right)\right]\:.
\end{equation}
Each of the Hamiltonians is similar to the $d=2$ dimensional Dirac field Hamiltonians obtained in (\ref{eq:dirac_ham_sym}). However, they are not exactly the same because of the presence of the $\kappa$-dependent global prefactor. The latter assures that the Hamiltonian for $\kappa=1$ is zero, so it can be explicitly removed from the sum.

To continue we  compute the equal time anticommutator of the magnetic modes  $\p_{\kappa \mu}^m$
\begin{equation}
\{\p_{\kappa \mu}^m(r),\p^{* m}_{\kappa ' \mu '}(r')\}=\left(\frac{\kappa-1}{\kappa}\right)\,\delta_{\kappa \kappa'}\delta_{\mu \mu'}\delta(r-r')\,,
\end{equation}
which motivates the following rescaling 
\begin{equation}
\p_{\kappa \mu}^m=\sqrt{\frac{\kappa-1}{\kappa}}R_{\kappa \mu}^m,\:\:\:\:\:\:\:\:\:\p_{\kappa \mu}^{*m}=\sqrt{\frac{\kappa-1}{\kappa}}R_{\kappa \mu}^{*m},
\end{equation}
so that the new redefined fields satisfy the canonical anticommutation relations (\ref{eq:commutF}) at equal times.
The rescaled Hamiltonian reads
\begin{equation}
 H_L =-\frac{i}{2}\sum_{\kappa \neq \pm 1,\,\mu }  \int_0^{\infty} dr \,\,\,\left(1-\frac{1}{\kappa}\right)\,\left[{R}_{-\kappa  \mu }^{m *} {R'}_{\kappa  \mu }^m - {R'}_{\kappa  \mu }^{m*} {R}_{-\kappa  \mu }^m +   \frac{\kappa}{r} \left( R_{-\kappa  \mu }^{m*}{R}_{\kappa  \mu }^m- R_{\kappa  \mu }^{m*}R_{-\kappa  \mu }^m \right)\right], \label{casi}
\end{equation}
where we have also integrated by parts the radial derivatives. The first contribution in (\ref{casi}) is just the reduced Dirac Hamiltonian (\ref{eq:dirac_ham_sym}) without the $\kappa =\pm 1$ modes. The second contribution is suppressed by a factor $\kappa^{-1}$ that changes the parity of the terms, so by summing over $\kappa$ it becomes a boundary term that can be discarded. 
Finally, writing the expression in terms of $\kappa>1$, the theory reduces to
\begin{equation}
H_L=\sum_{\kappa =2 }^{\infty}\sum_\mu \int_0^{\infty} dr\:\tilde{\Psi}^{\dagger}_{\kappa \mu}(r)\left(\alpha_r\:\overset{\leftrightarrow}{p}_{r}+\frac{\kappa}{r}\beta_r\right)\tilde{\Psi}_{\kappa \mu}(r),  \quad \tilde{\Psi}_{\kappa \mu}(r)=\begin{pmatrix}
R_{-\kappa \mu}(r) \\ R_{\kappa \mu} (r)
\end{pmatrix},
\label{eq:Ham_compact_notationrs}
\end{equation}
with the equal time anticommutation relations
\begin{equation}
\left\{\big(\tilde{\Psi}_{\kappa' \mu '}(r')\big)_i,\big(\tilde{\Psi}^{\dagger}_{\kappa \mu}(r)\big)_j\right\}=\delta_{ij} \delta_{\kappa \kappa'}\delta_{\mu \mu'}\delta(r-r').
\label{eq:commutRarita}
\end{equation}
In conclusion, using a mode decomposition in spherical harmonics we proved that the left Rarita-Schwinger Hamiltonian (\ref{eq:Ham_compact_notationrs}) can be written as the left Dirac Hamiltonian (\ref{eq:Ham_compact_notation}) excluding the $j=\frac{1}{2}$ contribution. In both cases the fields obey the same canonical anticommutation relations at equal times (\ref{eq:commutRarita}) and (\ref{eq:CommutDirac}). This is the main result of this subsection and will prove crucial for computing the universal coefficient of the EE for a Rarita-Schwinger field in Section \ref{subsec:logRarita}.
\subsection{Gauge fixing}
\label{subsec:Raritagauge}
To compute the  EE of any given region one must assign a von Neumann algebra of observables to such  region. In the Rarita-Schwinger theory, the gauge-invariant phase space is generated by the field strength (\ref{eq:fieldStrenght}). Therefore, to obtain the sphere EE from (\ref{eq:Ham_compact_notationrs}) we should describe how the gauge-dependant spinor field can be recovered from combinations of these gauge invariant operators localized on the sphere.

Note that the Rarita-Schwinger Hamiltonian  can be reduced to a tower of  Dirac Hamiltonians for any gauge choice. However, not all these choices allow us to write down the dynamical spinor $\vec{\psi}_L$. To be specific, the left-hand action of the gauge symmetry is given  by (\ref{gaugers}) as $\delta \vec{\psi}_L= \vec{\nabla} \chi_L$, so it is reasonable to expand out $\chi_L$ in the spinorial harmonics basis as
\begin{equation}
\chi_L (\vec{x})  =\sum_{\kappa \: \mu}\left(\frac{\chi_{\kappa \mu}(r)}{r}\right)\Omega_{\kappa  \mu }(\theta,\varphi). 
\end{equation}
From here we can see that the electric component of $\vec{\psi}_L$ can be easily set to zero since
\begin{equation}
\delta \vec{\psi}_L = \sum_{\kappa \: \mu} \left[\left(\chi'_{\kappa \mu}-\frac{\chi_{\kappa \mu}}{r}\right)\vec{\Omega}_{\kappa  \mu }^r  + \frac{\sqrt{\kappa(\kappa+1)}}{r^2} \chi_{\kappa \mu}\vec{\Omega}_{\kappa  \mu }^e \right]\:.
\end{equation}
Thus, by considering $\p^e_{\kappa \mu}(r)=0$ for all $\kappa$ and $\mu$ the spatial part of equation (\ref{lmd1}) reduces to
\begin{equation}
\vec{\psi_L} (\vec{x})  = \sum_{\:\kappa \: \mu}\left[\left(\frac{\p^r_{\kappa \mu}(r)}{r}\right)\vec{\Omega}^r_{\kappa  \mu }(\theta,\varphi) +\left(\frac{\p^m_{\kappa \mu}(r)}{r}\right)\vec{\Omega}^m_{\kappa  \mu }(\theta,\varphi) \right]\:.
\label{lmd12}
\end{equation}
The same procedure can be employed for  $\psi^\dagger_L$ and the  right-handed fields $\psi_R$  and $\psi^\dagger_R$ .

\
On the other hand, the electric  projection of the field strength can be written, for spatial Lorentz indices $a, b=1,2,3$, is 
\be
F^L_{e a} = \int d\Omega  \,[{\Omega^\dagger}^e_{\kappa \mu}]^{b}\, [F^L]_{a b} =   -\sum_{\kappa' \mu'}\int d\Omega  \,\Big([{\Omega^\dagger}^e_{\kappa \mu}]^{b} \partial_b [\psi^L_{\kappa' \mu'}]_a - [{\Omega^\dagger}^e_{\kappa \mu}]^{b} \partial_a [\psi^L_{\kappa' \mu'}]_b\Big)\,,
\ee
where the gauge choice (\ref{lmd12}) allows to eliminate all the derivatives of the spinor field present in the second term by means of
\bea 
\int d\Omega  \,[{\Omega^\dagger}^e_{\kappa \mu}]^{b} \partial_a [\psi^L_{\kappa' \mu'}]_b &=& \int d\Omega  \,\Big(\partial_a\big( [{\Omega^\dagger}^e_{\kappa \mu}]^{b}  [\psi^L_{\kappa' \mu'}]_b\big) - \big(\partial_a\, [{\Omega^\dagger}^e_{\kappa \mu}]^{b} \big) [\psi^L_{\kappa' \mu'}]_b \Big)\nonumber\\ &{}& \nonumber \\
&=& - \int d\Omega  \,\big(\partial_a\, [{\Omega^\dagger}^e_{\kappa \mu}]^{b} \big) [\psi^L_{\kappa' \mu'}]_b\:. \\ &{}& \nonumber 
\eea

Therefore, the electric projection of the field strength can be obtained from  the spinor field obeying the gauge condition (\ref{lmd12}) as
\be
F^L_{e a} = - \sum_{\kappa' \mu'}\int d\Omega  \,\Big([{\Omega^\dagger}^e_{\kappa \mu}]^{b} \partial_b [\psi^L_{\kappa' \mu'}]_a +\big( \partial_a\,[{\Omega^\dagger}^e_{\kappa \mu}]^{b} \big) [\psi^L_{\kappa' \mu'}]_b\,\Big)\:.
\ee
Although this relation is highly non-local, it only involves integrals and derivatives in the angular variables of the sphere at a fixed radius. Namely, it can be used to obtain an expression for  the gauge-fixed spinor in terms of the field strength that does not involve radial integrals that go outside the spherical region of interest.  

In addition, the identification of $\psi_L$ with $F_L$ given by (\ref{lmd12}) implies that they generate the same algebras with the same particle content. Thus this geometric gauge choice allows to describe pure spin $3/2$ algebras, without  propagating spin $1/2$ degrees of freedom\footnote{In the literature the propagating spin $1/2$ degrees of freedom are usually eliminated by the gauge choice $\gamma^\mu \psi_\mu = 0$. For this calculation, this choice is not convenient as it does not correspond to a clear geometric prescription.}, that can be assigned to spherical regions.

\subsection{Entanglement entropy and logarithmic coefficient}
\label{subsec:logRarita}
In conclusion, the EE of a massless Rarita-Schwinger field in $d=4$ in a spherical region is equivalent to the EE of a massless Dirac field excluding the contributions coming from the total angular momentum $j=\frac{1}{2}$ mode, i.e. with $\kappa=1$. The dynamics of such modes are described by the left-handed Hamiltonian
\begin{equation}
H_L^{\kappa=1}=\sum_{\mu=\pm \frac{1}{2}} \int_0^{\infty} dr\:\Psi^{\dagger}_{1\mu}(r)\left(\alpha_r\:\overset{\leftrightarrow}{p}_{r}+\frac{1}{r}\beta_r\right)\Psi_{1 \mu}(r)  \:.
\label{eq:Hamsubstract}
\end{equation}
These consist of a Dirac field on  the $d=2$ half-line $r>0$ with a $1/r$ mass term. To compute the EE on a finite segment $r\in(0,R)$, we can neglect the effects induced by the mass term because the UV divergent contributions come from high energy fluctuations near the entangling surface at $r=R$. In fact, its EE matches with the result for a free Dirac fermion in the half line \cite{Calabrese:2004eu}
\begin{equation}
    S_{\text{half-line}}=\frac{1}{6}\log\left(\frac{R}{\epsilon}\right),
    \label{EE1}
\end{equation}
for a UV cutoff $\epsilon$. 
We have explicitly verified this behavior numerically. Therefore,  the logarithmic coefficient for a massless Rarita-Schwinger field in $d=4$ is given by
\begin{equation}
c_{\text{log}}(3/2)=2\left(-\frac{11}{180}-2\left(\frac{1}{6}\right)\right)=-\frac{71}{90}, \label{7190}
\end{equation}
where $\frac{11}{180}$ is the chiral fermion contribution computed in Section \ref{subsec:latticeDirac}. The first factor of two takes into account both left/right chiralities, and the latter factor of two corresponds to the contribution from both modes with $\mu =\pm \frac{1}{2}$ whose entropy is given by (\ref{EE1}). Note that we have assumed that the EE for the Rarita-Schwinger field has the structure (\ref{eq:EntropyCFT}) of a CFT. This is because the theory exhibits conformal symmetry, as it can be explicitly inferred from the correlators of the gauge invariant field strength (refer to Appendix \ref{app:subsec:ScalevsConf} for a complete discussion). 
\begin{figure}[t]
    \centering
    \includegraphics[width=0.99\textwidth]{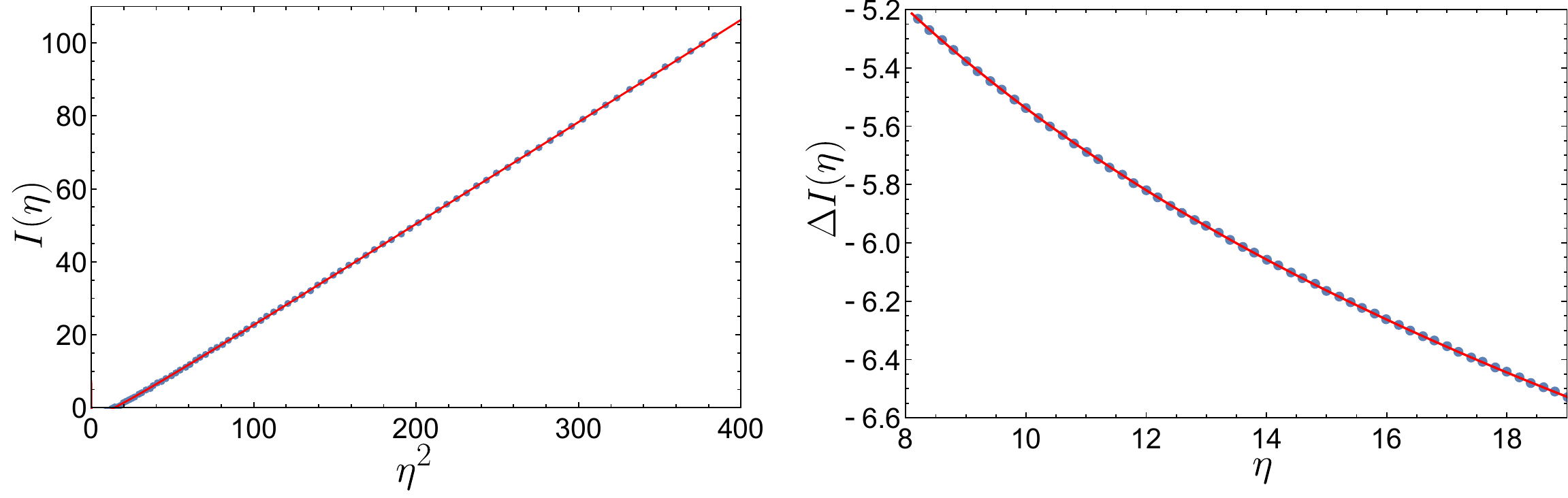}
    \caption{Left plot: Mutual information $I(\eta)$ for a Rarita-Schwinger field as a function of $\eta^2$, with $M=200$ physical sites. The  the fitting function is $f(\eta)=2\,c_2^{\text{fit}}\,\eta^2+2\,c^{\text{fit}}_{\text{log}}\,\log(\eta)+ c_0^{\text{fit}}$. Right plot: Mutual information with the area term subtracted, $\Delta I(\eta)=I(\eta)-2c_2^{\text{fit}}\,\eta^2$,  as a function of $\eta$. This exhibits the subleading logarithmic term appearing in the Rarita-Schwinger case $\Delta I(\eta)=2\,c^{\text{fit}}_{\text{log}}\,\log(\eta)+ c_0^{\text{fit}}$.}
    \label{fig:MutualRarita}
\end{figure}

We have numerically checked these results by applying the methods described in Section \ref{subsec:latticeDirac}, using the Hamiltonian (\ref{eq:Ham_compact_notationrs}). Indeed, the regularized entanglement entropy can be expanded as (\ref{eq:reg-ent}) and then numerically simulated, as shown in Fig. \ref{fig:MutualRarita}. The fitted coefficients in the continuum limit are
\be
s^{\infty}_2 (3/2)=0.14145 \,, \quad
c_{\text{log}}^{\infty}(3/2)=-0.79102\,,
\ee
which entail a $0.3 \%$ of relative error in compassion with the expected logarithmic coefficient $c_{\text{log}}(3/2)=-71/90=-0.788889$ (\ref{7190}). Again, the area term also coincides with the parallel plates approximation present in the literature \cite{Casini:2015dsg,Casini:2009sr} with $4.6\%$ of relative error. Note that the area term is not changed by subtracting the $\kappa=\pm1$ modes as these have a purely logarithmic contribution of the form (\ref{EE1}).

In light of these results, we can move forward to produce an extra argument for the conjecture presented in \cite{Dowker:2019zva}. Proceeding as in \cite{Benedetti:2019uej}, it is straightforward to think about a generalization for arbitrary free half-integer spin fields. Proposing that in a theory of half-integer spin $h$ the modes with momentum $\kappa<h-\frac{1}{2}$ cannot be excited we obtain 
\begin{equation}
c_{\text{log}}^{\text{fer}}(h)=2\left(-\frac{11}{180}-\frac{1}{6}\sum_{\kappa=0}^{h-1/2}(2\kappa)\right)=-\frac{7+60 h^2}{180}.
\label{subsec:highSpin}
\end{equation}
where the factor $1/6$ takes into account the logarithmic contributions coming from the low angular momentum modes that behave as (\ref{EE1}). Although this coincides with the proposal of \cite{Dowker:2019zva}, actual proof is still lacking in the literature. One possibility could be  to adapt the arguments in \cite{David:2020mls} for fermionic fields, or to perform a mode decomposition using more general tensor-spinor spherical harmonics. 


\section{Discussion}
\label{sec:concl}

We have computed the EE  for a spherical region for a free massless spin $3/2$ field. The result coincides with the EE associated with a spin $1/2$ field without the lowest angular momentum mode. In this case, the real-time approach clarifies that the entropy computed corresponds to that of the algebra of gauge invariant operators localized inside the sphere spanned by the spinor field strength. Thus, the computed entropy relates to pure spin 3/2 fields.

An issue we have avoided so far is the existence of ambiguities reaching the logarithmic coefficients (\ref{7190}) and (\ref{subsec:highSpin}). Both results have been obtained from the regularized EE, as they follow from the calculation performed in Section \ref{sec:Dirac} . Hence, the coefficients coincide with the universal terms appearing in the mutual information (\ref{midef}). The mutual information for separated regions is free of any ambiguities in the continuum limit \cite{Casini:2013rba,Casini:2019nmu}. Consequently, the aforementioned coefficients are well-defined quantities in the continuum field theory when interpreted as the universal coefficients of the mutual information.

The logarithmic coefficients (\ref{7190}) and (\ref{subsec:highSpin})  are subject to ambiguities when interpreted in terms of the EE (\ref{eq:EntropyCFT}). This can be made evident when computing the EE in the lattice. For every lattice theory, there are ambiguities regarding which operators  are included in the boundary of a given region. These elections represent particular choices of  algebras, which differ from each other because of the presence of specific centers. The elements of the corresponding center are labeled by the sectors of the theory and have an associated classical Shannon contribution to the EE that might change the universal coefficients \cite{Casini:2013rba}. These are known in  the literature as  "edge modes" contributions.

For instance, there used to be a well-known mismatch between the logarithmic coefficient for a free Maxwell field obtained from the conformal anomaly (\ref{eq:clogLowSpinLess1}), and the result obtained using the same real-time approach reported in the present article \cite{Casini:2015dsg},  or by studying thermodynamics in de Sitter spacetime \cite{Dowker:2010bu}, yielding (\ref{eq:clogLowSpinLessdif}). The edge modes contributions have been computed for the Maxwell field in \cite{Huang:2014pfa, Donnelly:2014fua, Donnelly:2015hxa, Soni:2016ogt,David:2022jfd} resolving the mismatch between the coefficients (\ref{eq:clogLowSpinLess1}) and  (\ref{eq:clogLowSpinLessdif}). Further, the gravitational edge modes  contribution for the linearized spin 2 theory has been computed in \cite{David:2022jfd}  in the context of a similar spherical wave decomposition as presented in the article. Such a procedure should be applicable to the Rarita-Schwinger field considering the sectors emerging from the constraints  (\ref{eq:constraints}).


In the continuum, general von Neumann algebras of observables do not contain a center, and the assignation of an algebra to spherical regions is expected to be unique. Nevertheless,  the ambiguities due to the presence of  sectors still persist, but they have a different physical origin \cite{Casini:2019nmu}. In the continuum, this difference emerges because $c_{\text{log}}$ behaves as an unprotected RG charge, in the sense that it depends on the UV completion of the theory \cite{Casini:2019nmu}. To be more specific, in the Maxwell theory,  the non-local IR correlations of some  Wilson and t'Hooft loops can be eliminated by adding a UV coupling to the charges, provided that the characteristic mass scale is larger than the regulating distance. The change in the contribution of these non-local operators near the boundary of the entangling region can be considered as an RG flow justification of the results obtained by adding ``edge modes'' when computing the EE. This kind of mismatch in the logarithmic coefficient is expected to arise not only in the Maxwell field but also in every theory that is not complete, namely, that exhibits sectors. 

The fact that ambiguities in the continuum are related to the UV completion of the theory complicates their interpretation for spin $h\geq 3/2$ fields. If we consider free higher spin theories as IR fixed points, several problems regarding possible RG flows arise. For instance, the free massless Rarita-Schwinger (minimally) coupled to an external electromagnetic field cannot be quantized while respecting all of the positivity constraints \cite{Adler:2015zha,Adler:2017lki}.  Indeed, following \cite{noether,noncompact} one should expect that these theories do not have a QFT-like well-defined RG flow, as they exhibit non-local sectors charged under the action of space-time symmetries. See  \cite{gravitonGS,Hinterbichler:2022agn} and \cite{Wang:2023iqt} for more explicit constructions of emergent IR generalized symmetries in the spin 2 and 3/2 cases, respectively.

These difficulties in the UV completion also obstruct a useful interpretation of the logarithmic coefficient. Theories with fields of spin $h\geq 3/2$  do not have a local gauge invariant stress-energy tensor, as it is prohibited by the Weinberg-Witten theorem \cite{Weinberg:1980kq}.  The lack of stress-energy tensor makes the notion of the conformal anomaly in curved backgrounds uncertain from the field theory standpoint and is therefore not reached by the A-theorem proof in \cite{Komargodski1,Komargodski2}.  On the other hand, the logarithmic term is still well-defined in the conformal field theory and even reached by the entropic A-theorem. In fact, the corresponding proof presented in \cite{Casini:2017vbe} requires only Poincaré invariance and strong subadditivity as assumptions, both of which are fulfilled in any free higher-spin theory.  Naively, this seems to imply that there still is an interpretation of the logarithmic coefficient as counting the degrees of freedom along  RG flows with free higher spin theories as IR fixed points, assuming that such flows exist.  However,  as we have suggested, this seems improbable.

In this context, the remaining possibility is that the RG flow is trivial, and the theory is always free. In such a case, there is of course a logarithmic term  in the EE, which coincides with the coefficient found in the mutual information, but does not flow. This indeed can be the interpretation of the coefficient for the free spin-$3/2$ field discussed in this article. Nevertheless, in this discussion, we are assuming the theory has to be UV completed in a QFT-like manner, so this argument does not necessarily have to hold for other types of  UV completions.  Perhaps the logarithmic coefficient computed here can be related to a non-trivial RG flow in such a context. This does seem a more reasonable approach, considering that the Rarita-Schwinger field appears in a wide variety of UV complete models of gravity (such as in supergravity models).

\section*{Acknowledgments}
We are grateful to Horacio Casini, Marina Huerta, Markus Luty, Mukund Rangamani, Gonzalo Torroba, Javier Magán, Pedro J. Martínez, and Andrew Waldron for useful discussions. We also like to thank Horacio Casini, Gonzalo Torroba, Mukund Rangamani for comments on the draft. We  thank Marina Huerta and Guido van der Velde for discussions about their upcoming work. We also thank Sophia Minnillo for proofreading the draft. The work of VB is supported by CONICET, Argentina. LD is supported by a Dean's Distinguished Graduate Fellowship from the College of Letters and Science of the University of California, Davis. 
\appendix

\section{Spherical harmonics}
\label{app:sphericalHarmonics}
\subsection{Spinor spherical harmonics}
\label{app:spinorh}
The spinor spherical harmonics are a complete basis of the space of spinor functions on the sphere  $S^2$. They are simultaneously eigenfunctions of the total angular momentum operator squared $\hat{J}^2$ and  the orbital angular momentum squared $\hat{L}^2$. The spinor spherical harmonics are useful for problems involving spinor fields on central potential (see for example \cite{Bjorken}). For the  convenience of the calculation, we will use the basis for spin $1/2$ spinors presented in \cite{Szmytkowski:2007mzc}, which defines the spinor spherical harmonics $\Omega_{\kappa \mu}$ in terms of the standard spherical harmonics\footnote{The scalar spherical harmonics $Y_{lm}(\theta,\varphi)$ involved in the definition are the usual ones, given in terms of the associated Legendre functions of the first kind $P_l^{(m)}(x)$ as $Y_{lm}(\theta,\varphi)=\sqrt{\frac{(2l+1)}{4\pi}\frac{(l-m)!}{(l+m)!}}P_l^{(m)}(\cos(\theta))e^{im\varphi}$.} $Y_{lm}$ as 
\begin{equation}
\Omega_{\kappa \mu}(\theta,\varphi)= \begin{pmatrix}
 \text{sgn}(-\kappa)\sqrt{\frac{\kappa+\frac{1}{2}-\mu}{2\kappa+1}} \: Y_{l,\mu - 1/2}(\theta,\varphi) \\ 
\sqrt{\frac{\kappa+\frac{1}{2}+\mu}{2\kappa+1}} \: Y_{l,\mu + 1/2}(\theta,\varphi) 
\end{pmatrix}
, \label{Om}
\end{equation}
where $0 \leq \theta \leq \pi$ and $0 \leq \varphi \leq 2\pi$ are the angular coordinates that parameterize $S^2$, and the quantum numbers $\kappa$ and $\mu$ are discretized as
\begin{equation}
    \kappa=\pm 1, \pm 2, \ldots \:,\:\:\:\:\mu = -|\kappa|+\frac{1}{2},-|\kappa|+\frac{3}{2},\ldots,|\kappa|-\frac{1}{2}.
\end{equation}
From here, the eigenvalues of the orbital angular momentum squared and the total angular momentum squared denoted by $l$ and $j$, respectively, can be recovered via
\begin{equation}
    l=\left|\kappa+\frac{1}{2}\right|-\frac{1}{2}=\begin{cases} 
    \kappa\:\:\:\:\:\:\:\:\:\:\:\:\:\:\: \text{for}\:\kappa>0\\
    -\kappa-1\:\:\:\: \text{for}\:\kappa<0
    \end{cases},\:\:\:\:\:\:\:\:\:\:\:\:j=|\kappa|-\frac{1}{2}\:.
    \label{eq:ljdef}
\end{equation}
\bigskip
In fact, the orbital angular momentum operator, given by $\vec{L} = -i\vec{r}\times \vec{\nabla}$, yields
\begin{equation}
    \hat{L}^2 \Omega_{\kappa \mu}(\theta, \phi) = \kappa(\kappa+1)\Omega_{\kappa \mu}(\theta, \phi)= l(l+1) \Omega_{\kappa \mu}(\theta, \phi),
\end{equation}
and the total angular momentum defined for spin $1/2$ as $\vec{J} = \vec{L} + \vec{\sigma}/2$ gives
\begin{equation}
    \hat{J}^2 \Omega_{\kappa \mu}(\theta, \phi) = (\kappa^2-1/4)\Omega_{\kappa \mu}(\theta, \phi)= j(j+1) \Omega_{\kappa \mu}(\theta, \phi).
\end{equation}
It is important to stress that the set of spherical harmonics (\ref{Om}) does not diagonalize the Dirac operator $-i\slashed{\nabla}$ on $S^2$. Notwithstanding, they can be written as a linear combination of functions that do so \cite{Abrikosov:2002jr}.

In addition, the orthonormality relations for spinor spherical harmonics are given by 
\begin{equation}
    \int_{S^2} d\Omega\:\Omega^{\dagger}_{\kappa'\mu'}(\theta,\varphi)\Omega_{\kappa,\mu}(\theta,\varphi)=\delta_{\kappa \kappa'}\delta_{\mu \mu'},
\end{equation}
which follows from the standard orthonormality conditions of scalar spherical harmonics.

\bigskip
An extended list of further properties can be found in \cite{Szmytkowski:2007mzc}. For the EE calculations, these are the relevant ones
\begin{align}
&(\vec{\sigma}\cdot \hat{r})\Omega_{\kappa  \mu }(\theta,\varphi)=-\Omega_{-\kappa  \mu }(\theta,\varphi)\:,\\
&(\vec{\sigma}\cdot \vec{L})\Omega_{\kappa  \mu }(\theta,\varphi)=-(\kappa+1)\Omega_{\kappa  \mu }(\theta,\varphi)\:, \\
& (\vec{\sigma}\cdot \vec{p})f(r)\Omega_{\kappa  \mu }(\theta,\varphi)=i\left(  \frac{\partial}{\partial r}+\frac{\kappa+1}{r} \right)f(r)\Omega_{-\kappa  \mu }(\theta,\varphi)\:,
\label{eq:sigmap}
\end{align}
where $\vec{p}=-i\vec{\nabla}$, $\vec{L}=\vec{r}\times \vec{p}$ and $\hat{r}=\frac{\vec{r}}{|\vec{r}|}$. Note that the equation (\ref{eq:sigmap}) follows from applying the differential operator $\vec{\sigma}\cdot \vec{p}$ in spherical coordinates
\begin{equation}
    \vec{\sigma}\cdot \vec{p}=(\vec{\sigma}\cdot \hat{r})\left(-i  \frac{\partial}{\partial r}+\frac{i \vec{\sigma} \cdot \vec{L}}{r} \right).
\end{equation}

\subsection{Vector-spinor spherical harmonics}
\label{app:vecspinorh}
Following the construction of the orthonormal basis of spinor spherical harmonics from scalar spherical harmonics, we define the vector-spinor spherical harmonics on $S^2$ from their vectorial counterparts as 
\begin{equation}
\vec{\Omega}^s_{\kappa \mu}(\theta,\varphi)= \begin{pmatrix}
 \text{sgn}(-\kappa)\sqrt{\frac{\kappa+\frac{1}{2}-\mu}{2\kappa+1}} \: \vec{Y}^s_{l,\mu - 1/2}(\theta,\varphi) \\ 
\sqrt{\frac{\kappa+\frac{1}{2}+\mu}{2\kappa+1}} \: \vec{Y}^s_{l,\mu + 1/2}(\theta,\varphi) 
\end{pmatrix} \,,\quad s=r,e,m
, \label{OmV}
\end{equation}
where $\vec{Y}^r_{lm}$, $\vec{Y}^e_{lm}$ and $\vec{Y}^m_{lm}$ are the radial, electric, and magnetic vector spherical harmonics conveniently defined as
\begin{align}
\vec{Y}^r_{lm} (\theta,\varphi) & = {Y}_{lm} (\theta,\varphi)  \,\hat{r} \,,\,\, \:\:\:\: \:\:\:\:\:\:\:\:\:\:\:\:\:\:\:\:l\geq 0\:\:\:\: \, -l \leq m \leq l  \\ & \nonumber \\
\vec{Y}^e_{lm} (\theta,\varphi) & = \frac{r \vec{\nabla}{Y}_{lm} (\theta,\varphi) }{\sqrt{l(l+1)}} \,,\,\,\:\:\:\: \:\:\:\:\:\:\:\:\:\:\:\:l\geq 1\:\:\:\: \, -l \leq m \leq l   \\ & \nonumber \\
\vec{Y}^m_{lm} (\theta,\varphi) & = \frac{\vec{r}\times \vec{\nabla}{Y}_{lm} (\theta,\varphi) }{\sqrt{l(l+1)}} \,,\,\, \:\:\:\: \:\:\:\:\:\:l\geq 1\:\:\:\: \, -l \leq m \leq l  \\  & \nonumber 
\end{align}
Note that  electric $\vec{Y}^e_{lm} $ and  magnetic $\vec{Y}^m_{lm} $ vector spherical harmonics are not defined for zero orbital angular momentum $l=0$. According to (\ref{eq:ljdef}), this corresponds to $\kappa=-1$. Thus, the vector-spinor spherical harmonics  $\vec{\Omega}^e_{-1 \mu}$ and $\vec{\Omega}^m_{-1 \mu}$ are not defined either.

\bigskip
Many of the properties of vector-spinor spherical harmonics are inherited from the properties of vector spherical harmonics, which can be found in \cite{Casini:2015dsg, Benedetti:2019uej, Thorne:1980ru, Compere:2017wrj}. For instance, we can check that ortonormality conditions hold as 
\begin{equation}
    \int_{S^2} d\Omega\:\vec{\Omega}^{\dagger s'}_{\kappa' \mu'}(\Omega) \cdot \vec{\Omega}^{s}_{\kappa \mu}(\Omega)=\delta_{ss'}\delta_{\kappa \kappa'}\delta_{\mu \mu'},
\end{equation}
which follows from the ortonormality of vector spherical harmonics given by
\begin{equation}
    \int_{S^2} d\Omega\:\vec{Y}_{l'm'}^{\dagger s'}(\Omega)\cdot\vec{Y}_{lm}^s(\Omega)=\delta_{s s'}\delta_{l l'}\delta_{m m'}.
\end{equation}
The directional and differential properties can be obtained in an analogous manner 
\begin{align}
 &\vec{\nabla } \cdot \vec{\Omega}^r_{\kappa \mu} =\frac{2}{r}\Omega_{\kappa \mu} \,,\qquad\qquad\quad\qquad\qquad\quad\,\,\,\,\, \hat{r} \cdot \vec{\Omega}^r_{\kappa \mu} = \Omega_{\kappa \mu}\:,\\
&\vec{\nabla } \cdot \vec{\Omega}^e_{\kappa \mu} = -\frac{\sqrt{\kappa(\kappa+1)}}{r}\Omega_{\kappa \mu}  \,,\qquad\qquad\qquad\hat{r} \cdot \vec{\Omega}^e_{\kappa \mu} =  0 \:,\\
&\vec{\nabla} \cdot \vec{\Omega}^m_{\kappa \mu} = 0 \,,\,\,\,\,\,\qquad\qquad\qquad\qquad\qquad\qquad \hat{r} \cdot \vec{\Omega}^m_{\kappa \mu} = 0 \:. \\ & \nonumber \\
&\vec{\nabla } \times \vec{\Omega}^r_{\kappa \mu} = -\frac{\sqrt{\kappa(\kappa+1)}}{r} \vec{\Omega}^m_{\kappa \mu} \,,\qquad\qquad\quad\,\,\, \hat{r} \times \vec{\Omega}^r_{\kappa \mu} = 0\:,\\
&\vec{\nabla } \times \vec{\Omega}^e_{\kappa \mu} = \frac{1}{r} \vec{\Omega}^m_{\kappa \mu} \,,\qquad\qquad\qquad\qquad\qquad\,\,\, \hat{r} \times \vec{\Omega}^e_{\kappa \mu} =  \vec{\Omega}^m_{\kappa \mu} \:,\\
&\vec{\nabla} \times \vec{\Omega}^m_{\kappa \mu} = -\frac{\sqrt{\kappa(\kappa+1)}}{r} \vec{\Omega}^r_{\kappa \mu} - \frac{1}{r} \vec{\Omega}^e_{\kappa \mu} \,,\qquad \hat{r} \times \vec{\Omega}^m_{\kappa \mu} =  -\vec{\Omega}^e_{\kappa \mu} \:.
\end{align}
 For EE calculations involving the Rarita-Schwinger field, the following properties were also employed
\begin{align}
  \vec{\sigma}\cdot \vec{\Omega}^r_{\kappa \mu}&=(\vec{\sigma}\cdot \hat{r})\Omega_{\kappa \mu}=-\Omega_{-\kappa  \mu }\;,\\ & \nonumber \\
\vec{\sigma}\cdot \vec{\Omega}^e_{\kappa \mu}&=\frac{ir(\vec{\sigma}\cdot\vec{p})}{\sqrt{\kappa(\kappa+1)}}\Omega_{\kappa \mu}=-\sqrt{\frac{\kappa+1}{\kappa}}\Omega_{-\kappa \mu}\:,\\ & \nonumber \\
\vec{\sigma}\cdot \vec{\Omega}^m_{\kappa \mu}&=\frac{i \vec{\sigma}\cdot \vec{L}}{\sqrt{\kappa(\kappa+1)}}\Omega_{\kappa \mu}=-i\sqrt{\frac{\kappa+1}{\kappa}}\Omega_{\kappa \mu}.
\end{align}
\section{Lattice fermions}
\label{app:lattice}
\subsection{Numerical simulations}
\label{app:numerical}
In this Appendix, we explain how to compute EE and mutual information numerically for a free Dirac field  in the half-line (\ref{eq:Ham_compact_notation}), following the procedure discussed in \cite{2003Peschel,Casini:2009sr}. First, we consider the model on a lattice with $r=na$ for $n \in \{1,2,\ldots,M\}$, infrared cutoff $M$ and lattice spacing $a$, which is set to $a=1$. For simplicity, we suppress the $(\kappa,\mu)$ indices and use the notation $\Psi_{\kappa \mu}(r)\to \psi_n$. Thus, the lattice version of the Hamiltonian (\ref{eq:Ham_compact_notation}) for each mode reads
\begin{equation}
    \hat{H}^{\kappa}=\sum_{n =1}^M\:\left[\frac{i}{2}\left(\psi_{n+1}^{\dagger}\alpha_r \psi_n-\psi_n^{\dagger} \alpha \psi_{n+1}\right)+\frac{\kappa}{n}\psi_n^{\dagger}\beta_r \psi_n \right]\equiv\sum_{n,m=1}^M \psi^{\dagger}_m\hat{H}^{\kappa}_{mn}\psi_n\:.
    \label{Hmat}
\end{equation}
We also impose the open chain boundary condition, i.e. $\psi_{M+1},\,\psi^{\dagger}_{M+1}=0$, since we do not expect correlations between $\psi_1$ and $\psi_M$ for large $M$. The lattice Hamiltonian  is described by a $2M \times 2M$ matrix  $\hat{H}^{\kappa}_{mn}$ because there are two degrees of freedom per site, with eigenvalues $\omega_q^{\pm}$ and eigenvectors $\vec{\Phi}_q^{\pm}$,
\begin{equation}
\hat{H}^{\kappa} \vec{\Phi}_q^{\pm}=\omega^{\pm}_q \vec{\Phi}^{\pm}_q,\:\:\:\:\:\:\:\:\:\:\: \vec{\Phi}_q^{\pm}=\begin{pmatrix}
\Phi^{\pm}_q(1)\:\cdots\:\Phi^{\pm}_q(l)\:\cdots\:\Phi^{\pm}_q(M)
\end{pmatrix}^T,\:\:\:\:\:\:\:\:\:\: q,l=1,\dots,M.
\end{equation}
In this context, $\pm$ means positive/negative energy, respectively, and $\Phi^{\pm}_q(k)$ without a vector symbol indicates that it is a two component spinor. The correlator involving fields in two sites $m,n$ can be computed as a sum over the outer product of negative energy spinors \cite{Eisler:2018ugn}
\begin{equation}
C_{mn}=\langle \psi_m^{\dagger }\psi_n  \rangle\equiv\langle G|\psi_m^{\dagger }\psi_n|G \rangle =\sum_{q=1}^M\Phi^{-\dagger }_q(m)\Phi^-_q(n).
\label{eq:Cmat}
\end{equation}
We only need $\omega_q^-=-\omega_q^+ \leq 0$ because the expectation values are computed with respect to the ground state $|G \rangle$, which is filled by all the negative energy states. For this construction, it is important that the eigenvectors $\vec{\Phi}_q^{\pm}$ are properly normalized to the unity. In models with translational invariance, it is possible to write down closed form expressions for $C_{mn}$ \cite{Daguerre:2020pte}. However, in the former case, $C_{mn}$ must be constructed by numerically diagonalizing $\hat{H}_{\kappa}$ and employing (\ref{eq:Cmat}).

 Since the theory is quadratic on the fields, the ground state $|G \rangle$ is Gaussian and the EE associated with any region  can be computed using the eigenvalues of  (\ref{eq:Cmat}).  If we restrict to a spatial subsystem $A$, we only need the eigenvalues of $C_A\equiv (C_{mn})_{m,n \in A}$, and the entropy can be computed as
\begin{equation}
 S(A)=-\text{Tr}\left[C_A\log(C_A)+(1-C_A)\log(1-C_A)\right].
 \label{entropyCV}
\end{equation}
For a finite line starting at the origin $(0,R)$, the corresponding discretization is achieved for $R=N+\frac{1}{2}$ and $m,n \in [1,N]$. For a finite line $(R_1,R_2)$ with $R_1<R_2$,  the corresponding discretization is achieved by $R_i=N_i+\frac{1}{2}$ and $m,n \in [N_1+1,N_2]$. Also, it is important to recall that for $d=2$ spacetime dimensions, because of the fermion doubling, the numerical results for the EE must be divided by $2$ to be consistent with the continuum limit. 

The mutual information for regions $A$ and $B$ can be defined in terms of the EE as $I(A,B)=S(A)+S(B)-S(A \cup B)$. Therefore, its numerical computation is straightforward and extends from the previous discussion. Particularly, it is useful for computing the regularized EE as discussed in Section \ref{subsec:latticeDirac}.
\subsection{Large angular momentum behavior}
\label{app:Largek}
In this Appendix, we prove that the leading order behavior of the EE for large angular momentum $\kappa$, for spherical region of radius $R=N+\frac{1}{2}$ on a lattice of size $M$, is independent of $M$ for $\kappa \gg M$ and is given by (\ref{eq:tailDirac}).

To begin with, we apply perturbation theory over the tridiagonal matrix $\hat{H}^{\kappa}_{mn}$ defined by the Dirac Hamiltonian on the lattice (\ref{Hmat}). Subsequently we split  $\hat{H}^{\kappa}_{mn}$ in a leading diagonal part $\hat{H}^{0}_{mn}$ of order $\mathcal{O}(\kappa)$ and in a subleading non diagonal one $ V_{mn}$ of order $\mathcal{O}(1)$. More precisely, we write
\begin{equation}
\hat{H}^{\kappa}_{mn}=\hat{H}^{0}_{mn} + g V_{mn}\,,\quad \hat{H}^{0}_{mn}= \frac{\kappa}{n}\,\beta_r\,\delta_{m\,n} \,,\quad V_{mn} = \frac{i}{2}\,\alpha_r\,\big( \delta_{m\,n+1} -\delta_{m+1\,n} \big),
\end{equation}
where  we have introduced the parameter $g$ to keep track of the perturbation order. The  eigenvalues and eigenvectors of the diagonal matrix $\hat{H}^{0}_{mn}$ are given by
\begin{equation}
\hat{H}^{0}_{mn}\vec{\Phi}_q^{\pm\, (0)}= \omega_q^{\pm\, (0)}\vec{\Phi}_q^{\pm\, (0)}\,,\quad \omega_q^{\pm\, (0)}= \pm \frac{\kappa}{q}\,,\quad  \vec{\Phi}_q^{\pm\, (0)}(n)=\frac{\delta_{qn}}{\sqrt{2}}\begin{pmatrix} \mp i \\ 1
\end{pmatrix} \:.
\end{equation}
Carrying out the perturbation up to second order gives the lattice Hamiltonian eigenvalues
\begin{equation}
    \omega_q^{\pm\, (2)}=\omega_q^{\pm\, (0)}\pm\frac{g^2}{4\kappa}\left[\left(\frac{1}{q}+\frac{1}{q-1}\right)^{-1}+\left(\frac{1}{q}+\frac{1}{q+1}\right)^{-1}\right].
\end{equation}
On the other hand, the eigenvectors are
\begin{equation}
\begin{split}
 \vec{\Phi}_q^{\pm\, (2)}& =  \vec{\Phi}_q^{\pm\, (0)} +\frac{g}{2\kappa}\left[b(q)\,\vec{\Phi}_{q+1}^{\mp\, (0)}-c(q)\,\vec{\Phi}_{q-1}^{\mp\, (0)} \right]\\
 &\:\:\:\:\:\:\:\:\:\:\:\:\:\:\:\:\:\:\:-\frac{g^2}{4\kappa^2}\left[\frac{1}{2}a(q) \vec{\Phi}_q^{\pm\, (0)} +d(q) \vec{\Phi}_{q+2}^{\pm\, (0)}+e(q) \vec{\Phi}_{q-2}^{\pm\, (0)}  \right]
 \label{ev1},
 \end{split}
\end{equation}
with coefficients
\begin{equation}
   a(q)=b^2(q)+c^2(q),\: \: \: \: \: \: \: \:  b(q)=\left({\frac{1}{q}+\frac{1}{q+1}}\right)^{-1},\: \: \: \: \: \: \: \: c(q)=\left({\frac{1}{q}+\frac{1}{q-1}}\right)^{-1},
\end{equation}
\begin{equation}
    d(q)=\left(\frac{1}{q}+\frac{1}{q+1}\right)^{-1}\left(\frac{1}{q}-\frac{1}{q+2}\right)^{-1},\: \: \: \: \: \: \: \: e(q)=\left(\frac{1}{q}+\frac{1}{q-1}\right)^{-1}\left(\frac{1}{q}-\frac{1}{q-2}\right)^{-1}.
\end{equation}
These eigenvectors are also normalized to the unity
\begin{equation}
    \langle \vec{\Phi}_q^{\pm\, (2)}| \vec{\Phi}_{q'}^{\pm\, (2)}\rangle=\delta_{q\:q'}+\mathcal{O}(g^3).
\end{equation}
In this set-up, we can compute the leading order correction to the correlation matrix $C_{mn}$ by simply replacing (\ref{ev1}) in the definition (\ref{eq:Cmat}). As a result, one obtains the matrix up to second order in $g$
\begin{equation}
    C_{n,n}=\frac{1}{2}\begin{pmatrix}  1 & -i \\ i & 1 
\end{pmatrix}-\frac{g^2}{4 \kappa^2}a(n)\begin{pmatrix}  0 & -i \\ i & 0
\end{pmatrix},
\end{equation}
\begin{equation}
    C_{n+1,n}=-C_{n,n+1}=\frac{g}{2\kappa}b(n)\begin{pmatrix}  0 & i \\ i & 0 
\end{pmatrix},
\end{equation}
\begin{equation}
    C_{n+2,n}=-C_{n,n+2}=-\frac{g^2}{8\kappa^2}\left[d(n)+e(n+2)-c(n+1)b(n+1)\right]\begin{pmatrix}  0 & -i \\ i & 0 
\end{pmatrix},
\end{equation}
with the remainder of the coefficients being zero.
Surprisingly, the entries of this matrix are independent of the total lattice size $M$. We have explicitly verified this phenomenon in our numerical simulations.

The EE of a spherical region of radius $R=N+\frac{1}{2}$ can be computed from the eigenvalues of $C_A=(C_{mn})_{m,n \in A}$ as in (\ref{entropyCV}). At zeroth order in $g$, the matrix $C_{A}$ is diagonal, with eigenvalues 0 and 1, and eigenvectors $\vec{\Phi}_q^{+\, (0)}$ and $\vec{\Phi}_q^{-\, (0)}$, respectively, yielding a net zero entropy $S(A)=0$. Corrections to the eigenvalues can be computed by applying degenerate perturbation theory to $C_A$. At first order in $g$, there are no corrections. However, at second order, the degenerate formalism is quite tedious to apply.
It is simpler to compute the corrections to the eigenvalues by examining the characteristic polynomial of $C_A$ at the lowest non trivial order in $g$. Rewriting the eigenvalues as
\begin{equation} \lambda=\lambda^{(0)}+g^2 \lambda^{(2)},
\end{equation}
with $\lambda^{(0)}=0,1\:$, it turns out that
\begin{equation}
\begin{split}
0&=\det{[C_A-\lambda \mathbb{I}_{2N }]}\\
&=f(\lambda^{(0)},\lambda^{(1)})+g^{2N}\left[\big(\lambda^{(2)}\big)^{N-1}\left(\lambda^{(2)} +(2\lambda^{(0)}-1) \frac{\xi(N)}{\kappa^2}\right)\right]+\mathcal{O}\big(g^{2(N+1)}\big),
\end{split}
\end{equation}
with
\begin{equation}
\xi(N)=\frac{N^2(N+1)^2}{4(2N+1)^2}.
\end{equation}
The function $f$ contains all of the terms up to $\mathcal{O}(g^{2(N-1)})$ and vanishes when $\lambda^{(0)}=0,1\:$ for all $\lambda^{(2)}$. Finally, the eigenvalues at second order are then given by $0$ and $1$ with multiplicity $N-1$, in addition to $\xi/\kappa^2$ and $(1-\xi/\kappa^2)$. Hence, the EE at first non trivial order for $\kappa \gg M$ 
\begin{equation}
\begin{split}
4\kappa S^{\kappa}_{\text{sphere}}(N,M)&=-4 \kappa\left[\frac{\xi}{\kappa^2}\log\left(\frac{\xi}{\kappa^2}\right)+\left(1-\frac{\xi}{\kappa^2}\right)\log\left(1-\frac{\xi}{\kappa^2}\right)\right]\\
&\sim 8\xi(N)\frac{\log(\kappa)}{\kappa}+4\xi(N)\big[1-\log(\xi(N))\big]\frac{1}{\kappa}+\ldots,
\end{split}
\label{eq:tailDiracw}
\end{equation}
where we have accounted for a factor of two due to the fermion doubling.
\section{Scale vs conformal invariance}
\label{app:subsec:ScalevsConf}

The free massless Rarita-Schwinger theory exhibits Poincaré and scale invariance; thus, in this appendix, we study whether such symmetry can be enhanced to include the full conformal group in $d=4$. On the one hand, we have stated that the theory describes massless particles of spin $3/2$. Therefore, the Weinberg-Witten theorem \cite{Weinberg:1980kq} precludes the existence of a local symmetric gauge invariant stress-energy tensor that can be integrated to produce the generators of the Poincaré group.  This failure can be considered an argument against the theory having conformal symmetry. In this context, the theory does not satisfy the usual CFT axioms appearing in the bootstrap literature, which assume the existence of such a current  \cite{Poland:2018epd,Simmons-Duffin:2016gjk}\footnote{Because of the lack of a well-defined stress-tensor, the  Rarita-Schwinger field evades many general results about the enhancement of the  Poincaré+scale-invariant symmetries \cite{Luty:2012ww,Dymarsky:2013pqa}, as it does not satisfy their assumptions.}. On the other hand, the lack of a well-defined stress-energy tensor does not imply the absence of conformal symmetry \cite{Kravchuk:2021kwe}. Indeed, this is the case for free higher integer spin theories  in $d=4$, which have been explicitly shown to have such symmetry \cite{Farnsworth:2021zgj,Longo:2018nyi}. 


Considering the spin-$3/2$ case, the corresponding free massless relativistic wave equation has been proven to be conformal \cite{Nakayama:2013is,Fushchich:1978hm}. Moreover,  all the free, massless, irreducible representations of the conformal group  have been  classified in \cite{Mack:1975je,Siegel:1988gd}, where the representations $(3/2,0)$ and $(0,3/2)$  of  $SO(4,2)$ are given by a spinor self/anti-self dual 2-forms fields satisfying the gamma-traceless condition. These are the same symmetries and equations of motion of the chiral/anti-chiral parts of $F_{\mu \nu}$ (\ref{rfl}). Further, the field strength $F_{\mu \nu}$ and its adjoint are the lowest dimensional operators appearing in the gauge invariant phase space, with their scaling dimension $\Delta =5/2$ saturating the unitarity bound \cite{Minwalla:1997ka}. 

In  this context, we will proceed  in the same vein as the proofs of conformal invariance presented for the Maxwell field \cite{El-Showk:2011xbs} and the linearized graviton field \cite{Farnsworth:2021zgj}. Specifically, we check that the two-point function of the spinor field strength $F_{\mu \nu}$ coincides with the correlation function of a spin-3/2 conformal primary, and that it transforms covariantly under conformal transformations. Moreover, because the theory is free, by Wick's theorem, all higher point correlators are defined solely in terms of two-point functions, which implies that the properties we proved extend to all of the correlators of the theory. In other words, the free massless Rarita-Schwinger theory  in $d=4$ describes a free conformal field theory with the field strength $F_{\mu \nu}$ as a gauge invariant conformal primary of spin $3/2$.

We now proceed to compute the two-point function to verify that $F_{\mu \nu}$ is a primary operator of the theory. The propagator of the vector-valued spinor field $\psi_\nu(x)$ can be obtained by properly inverting the quadratic action (\ref{eq:RaritaLagTot}) while taking into account the gauge redundancies \cite{freedman2012}
\begin{equation}
    \langle \psi_{\mu}(x)\bar{\psi}_{\nu}(y)\rangle=\frac{1}{4\pi^2}\left[\eta_{\mu \nu}\slashed{\partial}+\frac{1}{2}\gamma_{\mu}\slashed{\partial}\gamma_{\nu}\right]\frac{1}{(x-y)^{2}}+\text{ [gauge terms]},
\end{equation}
Therefore, the field strength and its adjoint yield by direct computation the correlator\footnote{The presence of $\gamma^5$ does not indicate a $O(4,2)$ invariance breaking to $SO(4,2)$. Indeed, considering the definition $\gamma^5= \frac{i}{4!}\epsilon_{\mu\nu  \rho \sigma}\gamma^{\mu} \gamma^{\nu}\gamma^{\rho} \gamma^{\sigma}$ we get $i\gamma^5 \epsilon^{\mu \nu \alpha \beta}=-\frac{1}{4!}\gamma^{\lambda} \gamma^{\sigma}\gamma^{\rho} \gamma^{\eta}\epsilon_{\lambda \sigma  \rho \eta}\epsilon^{\mu \nu \alpha \beta}=\frac{1}{4!}\gamma^{\lambda} \gamma^{\sigma}\gamma^{\rho} \gamma^{\eta} \eta_{\lambda \sigma  \rho \eta}^{\mu \nu \alpha \beta}$, where $\eta_{\lambda \sigma  \rho \eta}^{\mu \nu \alpha \beta}$ is the generalized Minkowski metric.}
\begin{align}
    \langle F_{\mu \nu}(x)\bar{F}_{\lambda \sigma}(y)\rangle &=-\frac{4}{\pi^2}\frac{\cancel{(x-y)}}{(x-y)^6}\Big[\mathcal{I}_{\mu \lambda}(x-y)\mathcal{I}_{\nu \sigma}(x-y)-(\mu \leftrightarrow \nu)\Big] \nonumber \\
    &-\frac{1}{\pi^2\,(x-y)^6}\Big[\mathcal{I}_{\mu \lambda}(x-y)\gamma_{\nu}\:\cancel{(x-y)}\:\gamma_{\sigma}+\mathcal{I}_{\nu \sigma}(x-y)\gamma_{\mu}\:\cancel{(x-y)}\gamma_{\lambda} -(\mu \leftrightarrow \nu)\Big] \nonumber\\
    &-\frac{\cancel{(x-y)}}{\pi^2\,(x-y)^6}i\gamma^5\tensor{\epsilon}{_{\mu \nu}^{\alpha \beta}} \Big[\mathcal{I}_{\alpha \lambda}(x-y)\mathcal{I}_{\beta \sigma}(x-y)-(\alpha \leftrightarrow \beta)\big]\:,  \label{eq:CorrFFbar}
\end{align}
where $\mathcal{I}_{\mu \nu}$ denotes the usual inversion tensor
\begin{equation}
    \mathcal{I}_{\mu \nu}(x-y)=\eta_{\mu \nu}-2\,\frac{(x-y)_{\mu}(x-y)_{\nu}}{(x-y)^2}.
    \label{eq:InveOp}
\end{equation}

 It is possible to check that (\ref{eq:CorrFFbar}) coincides with the two-point function of an irreducible spin-$\frac{3}{2}$ conformal primary using technology from a six-dimensional embedding space where conformal transformations are linearly realized \cite{Weinberg:2010fx,Weinberg:2012mz}. This technique is briefly reviewed in Appendix \ref{app:embeddingspace}, where we also give the expression in the six-dimensional space, which after being projected out to four dimensions reproduces (\ref{eq:CorrFFbar}).

In addition, we can check that (\ref{eq:CorrFFbar}) transforms covariantly under conformal transformations. This is self-evident for Poincaré or scale transformations; therefore, it remains to be proven for special conformal transformations. These transformations are tedious to implement, so another approach consists of demonstrating its covariance under spatial inversions \cite{Schreier:1971um,Rychkov:2016iqz}
\begin{equation}
    x'^{\mu}=-\frac{x^{\mu}}{x^2}\:,
\end{equation}
which implies covariance under special transformations, as these are compositions of inversions and translations. Indeed, we can compute
\begin{equation}
     \langle F^i_{\mu \nu}(x)\bar{F}^j_{\lambda \sigma}(y)\rangle={(x^2 y^2)^{-\frac{5}{2}}}\tensor{\mathcal{I}}{_{\mu}^{\alpha}}(x)\tensor{\mathcal{I}}{_{\nu}^{\beta}}(x)\tensor{\mathcal{I}}{_{\lambda}^{\rho}}(y)\tensor{\mathcal{I}}{_{\sigma}^{\delta}}(y)\: S^{ik}(x)S^{lj}(y)\langle F^k_{\alpha \beta}(x')\bar{F}^l_{\rho \delta}(y')\rangle \:,
     \label{eq:covarianceFF}
\end{equation}
with the fermionic inversion operator being defined as
\begin{equation}
    S^{ij}(x)=\frac{x^{\mu}\gamma_{\mu}^{ij}}{|x|}\:.
\end{equation}
The result (\ref{eq:covarianceFF}) follows from properties of the inversion tensor \cite{Ferrara:1972uq}. In particular, we consider that
\bea
    &{}&\tensor{\mathcal{I}}{_{\mu \nu}}(x)=    \tensor{\mathcal{I}}{_{\mu \nu}}(x'),\:\:\:\:\:\:\:\:\:\:\:\: \mathcal{I}_{\mu \nu}(x'-y')= \tensor{\mathcal{I}}{_\mu ^{\rho}}(x)\tensor{\mathcal{I}}{_\mu ^{\sigma}}(x)\mathcal{I}_{\rho \sigma}(x-y), \\
    &{}&\tensor{\mathcal{I}}{_\mu ^{\sigma}}(x)\tensor{\mathcal{I}}{_{\sigma \nu}}(x)=\eta_{\mu \nu},\:\:\:\:\:\:\:\tensor{\mathcal{I}}{_\mu ^{\alpha}}(x)\tensor{\mathcal{I}}{_\nu ^{\beta}}(x)\tensor{\mathcal{I}}{_\rho ^{\lambda}}(x)\tensor{\mathcal{I}}{_\delta^{\sigma}}(x)\tensor{\epsilon}{_{\alpha \beta}^{\rho \delta}} =-\tensor{\epsilon}{_{\mu \nu}^{\lambda \sigma}} ,
\eea
as well as the following relations for the fermionic inversion operator
\bea
    &{}&S^{ik}(x)S^{kl}(x'-y')S^{lj}(y)=S^{ij}(x-y)\:, \\
    &{}&\tensor{\mathcal{I}}{_{\mu} ^{\lambda}}(x)\tensor{\mathcal{I}}{_{\nu} ^{\sigma}}(y)S^{ik}(x)\:\gamma_{\lambda}^{kl}\:S^{lm}(x'-y')\:\gamma_{\sigma}^{mn}S^{nj}(y)=\gamma_{\mu}^{ip}S^{pq}(x-y)\gamma_{\nu}^{qj}\:.
\eea

These results demonstrate that the gauge invariant field strength correlators for the free massless Rarita-Schwinger field in $d=4$ have conformal symmetry. However, the theory does not have a local gauge invariant stress tensor, which prevents the construction of the generators of the conformal group. Conveniently,  the existence of operators, called twists, that implement the symmetry over a given space-time region is  secured by the weak version of Noether's theorem \cite{Buchholz:1985ii}.
The only requirement of the theorem is that the split property holds for every space-time region. This follows from two premises: the validity of the time slice axiom, and the fact that the degrees of freedom of the theory do not increase rapidly in the UV, namely, that there is no Hagerdon temperature. Both are expected to hold  for the free Rarita-Schwinger field on a Minkowski background.

For instance, twist operators for space-time translations can be constructed in the following manner.  Translational invariance of the theory provides a canonical stress tensor $\Theta_{\mu\nu}$ as a consequence of the usual Noether's theorem, 
\be
\Theta_{\mu\nu}=-\frac{1}{2}\Big[\epsilon_{ \mu \alpha\beta \gamma}\bar{\psi}^{\alpha}\gamma^5\gamma^{\beta}\partial_{\nu}\psi^{\gamma}- \eta_{\mu \nu}  \big(\epsilon_{ \alpha \beta \gamma \delta}\bar{\psi}^{\alpha}\gamma^5\gamma^{\beta}\partial^{\gamma}\psi^{\delta}\big)\Big]\:.
\ee
Such tensor is  not symmetric nor gauge invariant and the standard Belinfante procedure does not provide an improvement to make it gauge invariant \cite{Allcock:1976qp,freedman2012}. This can be interpreted as a direct consequence of the Weinberg-Witten theorem. Nevertheless, it can be used to provide a gauge invariant generator for spacetime translations  when integrated over the complete manifold \cite{freedman2012}. It can also be used to provide twist operators generating space-time translations in a given region if the correct boundary terms are added (see \cite{noether} for some examples).

\subsection{Embedding space}
\label{app:embeddingspace}
In this Appendix, we briefly review the six-dimensional embedding space formalism that gives a framework for computing correlation functions of primary fields in four-dimensional  CFTs. We mainly follow the conventions presented in \cite{Weinberg:2010fx,Weinberg:2012mz}. Furthermore, we present explicit expressions for terms in the embedding space whose projection to four dimensions reproduces the field strength correlator (\ref{eq:CorrFFbar}) of the free massless Rarita-Schwinger theory.

The basic idea of this method is that conformal transformations associated with the four dimensional group $SO(4,2)$ can be realized as linear transformations in a six-dimensional projective space defined via the hypercone condition
\begin{equation}
    \eta_{KL}X^K X^L=0,
\end{equation}
where for every non zero $\lambda$ the coordinates $X^K \sim \lambda X^K$ are identified. The capital indices $K,L,..=1,2,..6$ are Lorentz indices on the six-dimensional embedding space where the metric is given by\footnote{The signature in the four-dimensional sector $\eta_{\mu \nu}=(-+++)$ for $\mu,\nu=0,..,3$ is the opposite of the signature used throughout the article. Since tensorial expressions will not depend on the specific signature, in this Appendix, we opted to use the same conventions as \cite{Weinberg:2010fx,Weinberg:2012mz}.}
\begin{equation}
\eta_{11}=\eta_{22}=\eta_{33}=\eta_{55}=+1,\:\:\:\:\eta_{00}=\eta_{66}=-1\:.
\end{equation}
The action of four-dimensional conformal transformations coincides with the action of the Lorentz group in this six-dimensional embedding space, which is described by  
\begin{equation}
    X^{K}=\tensor{\Lambda}{^K_L}X^L \:\:\:\:\:,\:\:  \eta_{KL}\tensor{\Lambda}{^K_M}\tensor{\Lambda}{^L_N}=\eta_{MN} \:\:\:\:\:,\:\:  \det(\Lambda)=1\,.
\end{equation}
Further, the four dimensional coordinates $x^{\mu}$ can be projected out from the six dimensional ones $X^{\mu}$ by virtue of the transformation 
\begin{equation}
    x^{\mu}=\frac{X^{\mu}}{X^5+X^6}\:.
\end{equation}
As a warm-up, we consider primary tensor fields  belonging to the irreducible representations $(l,0)$ or $(0,l)$ for an integer $l$ in four dimensions. These are given by tensors $f_{\mu_1 \nu_1 ,\mu_2 \nu_2,...,\mu_l \nu_l}(x)$,
with $l$ pairs of skew-symmetric indices $[\mu_i \nu_i]$. Following the discussion in \cite{Weinberg:2010fx},  these tensor fields can be projected out from a tensor field $F_{K_1 L_1 ,K_2 L_2,...,K_l L_l}(X)$  with the same symmetries living in the six-dimensional space
\begin{equation} 
    f_{\mu_1 \nu_1 ,\mu_2 \nu_2,\dots,\mu_l \nu_l}(x)=(X^5+X^6)^{\Delta} e^{K_1}_{\mu_1}(x)e^{L_1}_{\nu_1}(x)\dots e^{K_l}_{\mu_l}(x)e^{L_l}_{\nu_l}(x) F_{K_1 L_1 ,K_2 L_2,...,K_l L_l}(X)\Big |_{\text{cone}},
\end{equation}
where $e_{\mu}^{K}(x)$ are the projection operators defined as
\begin{equation}
    e^{\mu}_{\nu}(x)=\delta^{\mu}_{\nu},\:\:\:\:\:e^{5}_{\mu}(x)=-x_{\mu},\:\:\:\:\:e^{6}_{\mu}(x)=x_{\mu}\:.
\end{equation}
The six dimensional terms involving $F_{K_1 L_1 ,K_2 L_2,...,K_l L_l}$ must be homogeneous under rescaling
\begin{equation}
     F_{K_1 L_1 ,K_2 L_2,...,K_l L_l}(\lambda X)= \lambda^{-\Delta} F_{K_1 L_1 ,K_2 L_2,...,K_l L_l}(X),
\end{equation}
with $\Delta$ as the scaling dimension of the four-dimensional conformal primary. In addition, they must satisfy the transversality condition for every free Lorentz index
\begin{equation}
    X^{K_1} F_{K_1 L_1 ,K_2 L_2,...,K_l L_l}( X)=0\:. \label{te}
\end{equation}
For instance, the correlator of a spin-1 conformal 2-form primary can be recovered from the corresponding correlator of the 2-form $F_{KL}(X)$ in the six-dimensional space. The most general two-point function is simply given by
\be 
    \langle F_{MN}(X)  F_{KL}(Y) \rangle =\frac{I_{MK}I_{NL}-I_{ML}I_{NK}}{(XY)^\Delta} ,\label{maxe}
\ee
where $I_{KL}$ denotes the tensor
\be 
I_{KL}(X,Y)=\eta_{KL}-\frac{X_L Y_K}{XY}\label{inve}\:,
\ee
which obeys the useful transversality properties
\be 
X^K I_{KL}= 0 \,, \quad Y^L I_{KL}= 0 \,,
\ee
thus implying that (\ref{maxe}) obeys the transversality condition (\ref{te}). Picking $\Delta=2$ saturates the spin 1 unitarity bound, and the four dimensional conformal correlator recovers the field theoretical result from the Maxwell field in $d=4$ \cite{El-Showk:2011xbs}.

The introduction of spin-1/2 fermions in the embedding space is more subtle. First, a set of gamma matrices that satisfy the six-dimensional Clifford algebra $\{\Gamma^K,\Gamma^L\}=2\eta^{KL}$ needs to be put in place. Those gamma matrices can explicitly be written as \cite{Weinberg:2010fx,Weinberg:2012mz} 
\begin{equation}
    \Gamma^\mu=\begin{pmatrix} 0 &  i \gamma^5 \gamma^\mu \\
        i \gamma^5 \gamma^\mu & 0 
    \end{pmatrix}\:,\quad \Gamma^5=\begin{pmatrix} 0 &   \gamma^5  \\
         \gamma^5 & 0 
    \end{pmatrix}\:,\quad \Gamma^6=\begin{pmatrix} 0 &  \mathbb{I}_4  \\
         -\mathbb{I}_4  & 0 
    \end{pmatrix}\:, \label{gammae}
\end{equation}
where $\gamma^\mu$ are four-dimensional gamma matrices that satisfy the four-dimensional Clifford algebra $\{\gamma^{\mu},\gamma^{\nu}\}=2\eta^{\mu \nu}$, and also generate $\gamma^5=-i\gamma^0\gamma^1\gamma^2\gamma^3$. As a result of this choice, a spinorial field $\Psi(X)$ in the embedding space can be  decomposed in its chiral parts by 
\begin{equation}
    \Psi(X)=\begin{pmatrix}
        \Psi_+(X)\\ \Psi_-(X)
    \end{pmatrix}\:.
\end{equation}
The corresponding spin-1/2 field in the four-dimensional space $\psi(x)$ is obtained as
\begin{equation}
    \psi(x)=(X^5+X^6)^{\Delta-\frac{1}{2}}\left[\left(\frac{1-\gamma^5}{2}\right)\Psi_+(X)+\left(\frac{1+\gamma^5}{2}\right)\Psi_-(X)\right]\Bigg |_{\text{cone}},
    \label{eq:spinorfield4d}
\end{equation}
provided that $ \Psi( X)$ obeys  
\begin{equation} 
    \Psi(\lambda X)=\lambda^{-\Delta+\frac{1}{2}}\Psi(X),\quad (\Gamma \cdot X)\Psi(X)=0\:.
\end{equation}
The two-point function of  spin-1/2 conformal primaries follows from the projection of the correlator
\be
    \langle \Psi(X)  \overline{\Psi}(Y) \rangle =\frac{(\Gamma\cdot X)(\Gamma\cdot Y)}{(XY)^{\Delta+\frac{1}{2}}} \:,\label{dirace}
\ee
which has the right scaling and transversality condition, considering that
\be 
(\Gamma\cdot X)^2 \Big|_{\text{cone}}=  \frac{1}{2} \{ \Gamma_A ,\Gamma_B\}X^A X^B\Big|_{\text{cone}} = 0\,.
\ee
Picking $\Delta=3/2$ saturates the unitarity bound for spin 1/2 fields, and the four dimensional projection of (\ref{dirace}) coincides with the two-point function of a free fermion field in $d=4$.

Finally, tensor-spinors in four dimensions transforming as $(l+\frac{1}{2},0)$ or $(0,l+\frac{1}{2})$ for an integer $l$ are described by $\mathcal{F}_{\mu_1 \nu_1 ,\mu_2 \nu_2,\dots,\mu_l \nu_l}(x)$. Indeed, they can be represented in the embedding space by
\begin{equation}
    \mathcal{F}_{K_1 L_1 ,K_2 L_2,...,K_l L_l}(X)=\begin{pmatrix}
        \mathcal{F}^+_{K_1 L_1 ,K_2 L_2,...,K_l L_l}(X)\\ \mathcal{F}^-_{K_1 L_1 ,K_2 L_2,...,K_l L_l}(X)
    \end{pmatrix}\:.
\end{equation}
The corresponding  reduction to the four-dimensional space is thus
\begin{equation}
\begin{split}
    &\mathcal{F}_{\mu_1 \nu_1 ,\mu_2 \nu_2,\dots,\mu_l \nu_l}(x)=(X^5+X^6)^{\Delta-\frac{1}{2}}e^{K_1}_{\mu_1}(x)e^{L_1}_{\nu_1}(x)\dots e^{K_l}_{\mu_l}(x)e^{L_l}_{\nu_l}(x)\times\\
    &\:\:\:\:\:\:\:\:\:\:\:\:\:\:\:\:\:\:\:\:\times \Bigg[\left(\frac{1-\gamma^5}{2}\right) \mathcal{F}^+_{K_1 L_1 ,K_2 L_2,...,K_l L_l}(X)+\left(\frac{1+\gamma^5}{2}\right)\mathcal{F}^-_{K_1 L_1 ,K_2 L_2,...,K_l L_l}(X)\Bigg]\Bigg |_{\text{cone}},
\end{split}
\end{equation}
where $\Delta$ is the scaling dimension of the conformal primary in four dimensions, and
\begin{equation}
     \mathcal{F}_{K_1 L_1 ,K_2 L_2,...,K_l L_l}(\lambda X)=\lambda^{-\Delta+\frac{1}{2}}\mathcal{F}_{K_1 L_1 ,K_2 L_2,...,K_l L_l}( X), 
\end{equation}
along with the constraints
\begin{equation}
    X^{K_i}\mathcal{F}_{K_1 L_1 ,...,K_i L_i,...K_l L_l}(X)=0\,,\quad(\Gamma \cdot X)\mathcal{F}_{K_1 L_1 ,K_2 L_2,...,K_l L_l}( X)=0\:. \; \label{trae}
\end{equation}
For these higher-spin fields, extra irreducibility conditions must be imposed on $\mathcal{F}$. This is the double gamma-traceless requirement $\gamma^{\mu_1}\gamma^{\nu_1}\mathcal{F}_{\mu_1 \nu_1 ,\mu_2 \nu_2,\dots,\mu_l \nu_l}(x)=0$ that eliminates degrees of freedom that do not correspond to spin $l+\frac{1}{2}$ fields. This  can be also imposed in the embedding space by means of $\Gamma^{K_1}\Gamma^{L_1}\mathcal{F}_{K_1 L_1 ,K_2 L_2,...,K_l L_l}(X)=0$ \cite{Weinberg:2012mz}. However, it is simpler to impose this condition in the four-dimensional space after reduction.

When considering the case of interest, a spinor 2-form field $\mathcal{F}_{KL}$, we found three terms in the embedding space that  independently satisfy the transversality constraints (\ref{trae}). These are given by combinations of the tensor (\ref{inve}) and the gamma matrices (\ref{gammae}) as
\bea 
    &{}&\langle \mathcal{F}_{MN}(X)  \bar{\mathcal{F}}_{KL}(Y) \rangle =\frac{(\Gamma\cdot X)}{(XY)^{\Delta+\frac{1}{2}}}\Big[a_1 \,(I_{MK}I_{NL}-I_{ML}I_{NK} )+\nonumber  \\
    &{}& \qquad\qquad +  a_2(I_{MK}\Gamma_N \Gamma_L-I_{ML}\Gamma_N \Gamma_K+I_{NL}\Gamma_M \Gamma_K -I_{NK}\Gamma_M \Gamma_L) + \label{rse}\\
     &{}& \qquad\qquad + a_3\,\Gamma^A \Gamma^B \Gamma^C \Gamma^D\,(I_{MA}I_{NB}-I_{MB}I_{NA} )(I_{CK}I_{DL}-I_{CL}I_{DK} )\Big](\Gamma\cdot Y)\:,\nonumber 
\eea
for a priori arbitrary constants $a_1,\,a_2$ and $a_3$. The first term, proportional to  $a_1$, and the second term,  proportional to $a_2$, respectively, reduce to the first and second line of (\ref{eq:CorrFFbar}). The third term, proportional to $a_3$, reduces to a linear combination of the three lines of (\ref{eq:CorrFFbar}). Thus, they represent a set of three independent terms that  can recover the Rarita-Schwinger correlator in four dimensions. The gamma tracelesness condition $\gamma^{\mu}\langle F_{\mu \nu}(x)\bar{F}_{\lambda \sigma}(y)\rangle=0$ and the divergenceless condition $\partial^{\mu}\langle F_{\mu \nu}(x)\bar{F}_{\lambda \sigma}(y)\rangle=0$ uniquely fix the relative weight of each coefficient $a_i$ once the correlator (\ref{rse}) is projected out to four dimensions. The final result coincides with the field strength correlator (\ref{eq:CorrFFbar}) of the Rarita-Schwinger theory for the proper scaling dimension $\Delta=5/2$.

\

\bibliography{Raritabib}{}
\bibliographystyle{utphys}

\end{document}